\documentclass[12pt]{article}

\usepackage{amssymb,amsmath}

\usepackage{graphicx}
\DeclareGraphicsExtensions{.pdf,.png,.jpg}

\renewcommand{\theequation}{\arabic{section}.\arabic{equation}}
 \catcode`\@=11 \@addtoreset{equation}{section}\catcode`\@=12
\newcommand{\R}{ {\mathbb R} }

\begin{document}

 \begin{center}

 \large \bf Stable  exponential cosmological  solutions with three different Hubble-like parameters in EGB  model  with a $\Lambda$-term 
  \end{center}

 \vspace{0.3truecm}

 \begin{center}

   K. K. Ernazarov$^{1}$,  V. D. Ivashchuk$^{1,2}$

\vspace{0.3truecm}

  \it $^{1}$ 
    Institute of Gravitation and Cosmology, \\
    Peoples' Friendship University of Russia (RUDN University), \\
    6 Miklukho-Maklaya Street,  Moscow, 117198, Russian Federation, \\ 
 
  \it $^{2}$ Center for Gravitation and Fundamental Metrology,  VNIIMS, \\
  46 Ozyornaya Street, Moscow, 119361,  Russian Federation.

\end{center}

\begin{abstract}

We consider a $D$-dimensional  Einstein-Gauss-Bonnet model with a cosmological term $\Lambda$  and two non-zero
constants: $\alpha_1$ and $\alpha_2$. We   restrict the metrics to be diagonal ones and study  a class  of solutions with  exponential time dependence of three scale factors, governed by three non-coinciding Hubble-like parameters:
$H \neq 0$, $h_1$ and $h_2$, obeying  $m H + k_1 h_1 + k_2 h_2 \neq 0$ and corresponding to factor spaces of
dimensions $m > 1$, $k_1 > 1$ and $k_2 > 1$, respectively ($D = 1 + m + k_1 + k_2$). We analyse two cases: 
i) $m  <  k_1 < k_2$ and ii) $1< k_1 = k_2 = k$, $k \neq m$. We show that in both cases the solutions  exist  if 
$\alpha = \alpha_2 / \alpha_1 > 0$ and $\alpha \Lambda > 0$ satisfies certain  restrictions, e.g. upper and lower
bounds. In case ii) explicit relations for exact solutions are found. In both cases the subclasses of stable
and non-stable solutions are singled out. For $m > 3$ the case i) contains a subclass of solutions  describing
an exponential expansion of  $3$-dimensional  subspace with Hubble parameter $H > 0$ and zero variation of the effective gravitational constant $G$. The case $H = 0$ is also considered.

\end{abstract}


\section{Introduction}

In this paper we consider $D$-dimensional  Einstein-Gauss-Bonnet (EGB) model with a $\Lambda$-term. 
To some extent this model is unique among the other  higher-dimensional  extensions of 
General Relativity (GR) with second  order in curvature terms. The reason is  the following one: 
the equations of motion for this model are of the second order (in derivatives) like it takes place 
in the  Einstein gravity. It is well known that  the so-called Gauss-Bonnet term appeared in 
(super)string theory as a first order correction  (in $\alpha'$) to the  (super)string effective action (e.g. heterotic one)  \cite{Zwiebach}-\cite{GW}.   
 
 Currently, EGB gravitational model in diverse dimensions  and  its modifications,  
 see   \cite{Ishihara}-\cite{NOO-19} and refs. therein,  are rather popular objects for studying in  cosmology. 
 They are used for possible explanation  of  accelerating  expansion of the Universe 
 (i.e. solving the dark energy problem), which follow from supernova (type Ia) observational data 
 \cite{Riess,Perl,Kowalski}. One may  expect that the second order   form of the equations of motion for these models will lead us to solutions which are in some sense close to those coming from GR and its higher dimensional extensions 
 (e.g. avoiding the ghosts branches  at least).
  
 The $D$-dimensional EGB model is a particular case of the Lovelock model \cite{Lovelock}. 
 The equations of motion for the Lovelock model have also at most second order derivatives of the metric 
 (as it takes place in GR). We note that at present there exist several modifications of Einstein and EGB actions which correspond to $F(R)$, $R + f({\cal G})$, $f({R, \cal G})$ theories (e.g. for $D=4$), where $R$ is scalar curvature and  ${\cal G}$ is  Gauss-Bonnet term. These modifications are under intensive studying  devoted to cosmological, astrophysical and other applications, see \cite{NOO-17}-\cite{NOO-19} and references therein.

In this paper we restrict ourselves to diagonal  metrics  and study (mainly) a class of cosmological solutions with  exponential time dependence of three scale factors, governed by three non-coinciding Hubble-like parameters: $H \neq 0$, $h_1$ and $h_2$, corresponding to factor spaces of dimensions $m > 1$, $k_1 > 1$ and $k_2 > 1$, respectively, with a restriction imposed: $S_1 = m H + k_1 h_1 + k_2 h_2 \neq 0$,  and $D = 1 + m + k_1 + k_2$. This restriction   forbids the solutions with constant volume factor. We note that in generic anisotropic case with Hubble-like parameters $h_1, \dots, h_n$  obeying  $S_1 = \sum_{i=1}^n h_i \neq 0$ ($n = D-1$) the number of different real numbers among  $h_1, \dots, h_n$  should not exceed $3$ \cite{Ivas-16}.

Here we study  two cases: i) $m <  k_1 < k_2$ and ii) $1< k_1 = k_2 = k$, $k \neq m$. We show that in both cases the solutions  exist  only if $\alpha = \alpha_2 / \alpha_1 > 0$, $ \Lambda > 0$ and  $\Lambda$ obeys certain restrictions, e.g. inequalities of the form:  $0 < \lambda_{*}(m,k_1,k_2) < \Lambda \alpha <  \lambda_{**}(m,k_1,k_2)$. We note that in superstring inspired models $\alpha$ is positive and corresponds to Regge slope parameter  $\alpha'$ which is inverse proportional to the tension of the (super)string;  non-zero $ \Lambda$-terms appear for non-critical superstrings.

The solutions under consideration are reduced  to solutions of polynomial master equation  of fourth order or less, which may be solved in radicals for all $m > 1$, $k_1 > 1$ and $k_2 > 1$. In the case ii) $1< k_1 = k_2 = k$, $k \neq m$ we present explicit exact solutions for Hubble-like parameters.  Here we use our previous results from refs. \cite{ErIvKob-16,Ivas-16} in studying the stability of the solutions under consideration. In Section 5 we single out (for both cases i) and ii)) the subclasses of stable and non-stable solutions. In  Section 6  we present as an example  a subclass of solutions (for the case i)) describing an exponential expansion of  $3$-dimensional subspace with Hubble parameter $H > 0$ and zero variation of the effective gravitational constant $G$ (in Jordan frame) which was obtained in Ref. \cite{ErIv-17-2} for   fixed value of  $ \Lambda$ (depending upon  $ m, k_1, k_2$ and $\alpha > 0$). 

We note that earlier Ref. \cite{IvKob-18-2}  was dealing with exponential cosmological solutions    in the EGB model (with a $ \Lambda$-term) with two non-coinciding Hubble-like parameters   $H > 0$ and $h$ obeying $S_1 = m H + l h_1 \neq 0$ and corresponding to    $m$- and $l$-dimensional factor spaces ($m > 2$, $l>2$). In this case  there were two sets of solutions obeying: a) $\alpha > 0$, $ \Lambda < \alpha^{-1} \lambda_{+}(m,l)$ and  b) $\alpha < 0$, $  \Lambda > |\alpha|^{-1}  \lambda_{-}(m,l)$, with $\lambda_{\pm}(m,l) > 0$. Thus, the case of two (non-coinciding) Hubble-like parameters from Ref. \cite{IvKob-18-2}   drastically differs from   the case of three (non-coinciding) Hubble-like parameters which is studied in this paper.

\section{The cosmological model}

The action of the model reads
\begin{equation}
  S =  \int_{M} d^{D}z \sqrt{|g|} \{ \alpha_1 (R[g] - 2 \Lambda) +
              \alpha_2 {\cal L}_2[g] \},
 \label{2.0}
\end{equation}
where $g = g_{MN} dz^{M} \otimes dz^{N}$ is the metric defined on
the manifold $M$, ${\dim M} = D$, $|g| = |\det (g_{MN})|$, $\Lambda$ is
the cosmological term, $R[g]$ is scalar curvature,
$${\cal L}_2[g] = R_{MNPQ} R^{MNPQ} - 4 R_{MN} R^{MN} +R^2$$
is the standard Gauss-Bonnet term and  $\alpha_1$, $\alpha_2$ are
nonzero constants.

We consider the manifold
\begin{equation}
   M = \R  \times   M_1 \times \ldots \times M_n 
   \label{2.1}
\end{equation}
with the metric
\begin{equation}
   g= - d t \otimes d t  +
      \sum_{i=1}^{n} B_i e^{2v^i t} dy^i \otimes dy^i,
  \label{2.2}
\end{equation}
where   $B_i > 0$ are arbitrary constants, $i = 1, \dots, n$, and
$M_1, \dots,  M_n$  are one-dimensional manifolds (either $\R$ or $S^1$)
and $n > 3$.

The equations of motion for the action (\ref{2.0}) 
give us the set of  polynomial equations \cite{ErIvKob-16}
\begin{eqnarray}
  E = G_{ij} v^i v^j + 2 \Lambda
  - \alpha   G_{ijkl} v^i v^j v^k v^l = 0,  \label{2.3} \\
   Y_i =  \left[ 2   G_{ij} v^j
    - \frac{4}{3} \alpha  G_{ijkl}  v^j v^k v^l \right] \sum_{i=1}^n v^i 
    - \frac{2}{3}   G_{ij} v^i v^j  +  \frac{8}{3} \Lambda = 0,
   \label{2.4}
\end{eqnarray}
$i = 1,\ldots, n$, where  $\alpha = \alpha_2/\alpha_1$. Here
\begin{equation}
G_{ij} = \delta_{ij} -1, \qquad   G_{ijkl}  = G_{ij} G_{ik} G_{il} G_{jk} G_{jl} G_{kl}
\label{2.4G}
\end{equation}
are, respectively, the components of two  metrics on  $\R^{n}$ \cite{Iv-09,Iv-10}. 
The first one is a 2-metric and the second one is a Finslerian 4-metric.
For $n > 3$ we get a set of forth-order polynomial  equations.

We note that for $\Lambda =0$ and $n > 3$ the set of equations (\ref{2.3}) 
and (\ref{2.4}) has an isotropic solution $v^1 = \cdots = v^n = H$ only 
if $\alpha  < 0$ \cite{Iv-09,Iv-10}.
This solution was generalized in \cite{ChPavTop} to the case $\Lambda \neq 0$.

It was shown in \cite{Iv-09,Iv-10} that there are no more than
three different  numbers among  $v^1,\dots ,v^n$ when $\Lambda =0$. This is valid also
for  $\Lambda \neq 0$ if $\sum_{i = 1}^{n} v^i \neq 0$  \cite{Ivas-16}.

%\section{Cosmological solutions}

Here we consider a class of solutions to the set of equations (\ref{2.3}), 
(\ref{2.4}) of the following form:
\begin{equation}
  \label{3.1}
   v =(\overbrace{H, \ldots, H}^{m}, 
   \overbrace{h_1, \ldots, h_1}^{k_1}, \overbrace{h_2, \ldots, h_2}^{k_2}),
\end{equation}
where $H$ is the Hubble-like parameter corresponding  
to an $m$-dimensional factor space with $m > 1$,  $h_1$ is the Hubble-like parameter 
corresponding to an $k_1$-dimensional factor space with $k_1 > 1$ and $h_2$ is the Hubble-like parameter corresponding to an $k_2$-dimensional factor space with $k_2 > 1$. 
In Section 6 we split the $m$-dimensional  factor space for $m > 3$ into the  product of two subspaces of dimensions $3$ and $m-3$, respectively. The first one is identified with ``our'' $3d$ space while the second one is considered as 
a subspace of $(m-3 + k_1 + k_2)$-dimensional internal space.
 
 {\bf Remark.} {\em For $H > 0$ ``our'' 3d space expands
 (isotropically) with Hubble parameter $H$ and the $(m - 3)$-dimensional 
 part of internal space ($m > 3$) expands (isotropically)
 with the same Hubble parameter H too. Moreover, 
 we may deal with Hubble-like parameters decribing the internal subspaces 
 which  obey $h_1 > H$ or $h_2 > H$ (see Section 6).
 To avoid possible questions with the separation of  subspaces, we consider for physical applications 
 (in our epoch) the internal space to be compact, i.e. we put in  (\ref{2.1})
  $M_4 = \dots  = M_n  = S^1$ and  we  set the internal scale factors corresponding to the
 present time $t_0$: $a_ j (t_0) = (B_k)^{1/2}  exp(v^ j t_0)$, $k = 4, \dots , n,$ (see
 (\ref{2.2})) to be small enough in comparison with the scale factor
 of ``our'' space for $t = t_0$: $a(t_0) = B^{1/2}  exp(H t_0)$, where
 $B_1 = B_2 = B_3 = B > 0$.}
 
 We consider the ansatz (\ref{3.1}) with three Hubble-like parameters $H$, $h_1$ and $h_2$ 
which obey the following restrictions:
   \begin{equation}
     H \neq h_1, \quad  H \neq h_2, \quad 
     h_1 \neq h_2, \quad S_1 = m H + k_1 h_1 + k_2 h_2 \neq 0.
   \label{3.3}
   \end{equation}

 In Ref. \cite{ErIv-17-2}  the set of $(n + 1)$ polynomial equations  
 (\ref{2.3}), (\ref{2.4}) under ansatz  
 (\ref{3.1}) and restrictions (\ref{3.3}) imposed  was reduced to a set  of three polynomial equations 
 (of fourth, second and first orders, respectively)
    
    \begin{eqnarray}
          E =0,   \label{3.4E} \\
          Q =  - \frac{1}{2 \alpha}, \label{3.4Q} \\
          L = H + h_1 + h_2 - S_1 = 0.  \label{3.4L}
     \end{eqnarray}
   where  $E$ is defined in (\ref{2.3}) and 
   \begin{equation}
        Q = Q_{h_1 h_2} =  S_1^2 - S_2 - 2 S_1 (h_1 + h_2) + 2 (h_1^2 + h_1 h_2 + h_2^2),
                   \label{3.5}
        \end{equation}
   where here and in what follows 
        \begin{equation}
       S_k = \sum_{i =1}^n (v^i)^k.
       \label{3.5a}
        \end{equation}   
  This reduction is a special case of a more general   prescription 
  (Chirkov-Pavluchenko-Toporensky trick) from  Ref. \cite{ChPavTop1}. 
 
 Moreover, it was shown in Ref. \cite{ErIv-17-2} that the following relations
 take place
  \begin{equation}
   Q_{h_i h_j} =  S_1^2 - S_2 - 2 S_1 (h_i + h_j) + 2 (h_i^2 + h_i h_j + h_j^2)
      = - \frac{1}{2 \alpha},
       \label{3.5b}
   \end{equation}
  where $i \neq j$; $i, j = 0,1,2$ and $h_0 = H$.       

Due to (\ref{3.3}) the case $H = h_1 = h_2 = 0$ is excluded.  
First, we put
\begin{equation}
  \label{3.2a}
   H \neq 0. 
\end{equation}
          
      Let us denote 
       \begin{equation}
         x_1 = h_1/H, \qquad     x_2 = h_2/H.      
                  \label{3.8}
       \end{equation}        
    Then restrictions (\ref{3.3}) read 
      \begin{equation}
        x_1 \neq 1, \quad  x_2 \neq 1, \quad  x_1 \neq x_2, 
        \quad  m  + k_1 x_1 + k_2 x_2 \neq 0.
      \label{3.3a}
      \end{equation}

   Equation  (\ref{3.4L}) in $x$-variables reads
   \begin{equation}
    m -1  + (k_1 - 1) x_1 + (k_2 -1) x_2 = 0.  
   \label{3.15}
   \end{equation} 
   
   Here we should exclude from our consideration the case
    \begin{equation}
      \label{3.1a}
      m = k_1 = k_2.
    \end{equation}
       Indeed,   for  $m = k_1 = k_2 >1 $ we get from restriction (\ref{3.3a}): 
    $1  +  x_1 + x_2 \neq 0$,
   while (\ref{3.15}) gives us the relation $1  +  x_1 + x_2 = 0$, 
   which is  incompatible with the previous one.

    We get from (\ref{3.4Q}) and (\ref{3.5}) that 
            
     \begin{equation}
          2 \alpha {\cal P} H^2   = - 1,       \label{3.9}
      \end{equation} 
  where
     \begin{eqnarray}
     {\cal P}   = {\cal P} (x_1,x_2) = {\cal P} (x_1,x_2,m, k_1, k_2) = 
     \nonumber \\    
     (m + k_1 x_1 + k_2 x_2 )^2  - (m + k_1 x_1^2 + k_2 x_2^2) 
     \nonumber \\
     - 2 (m + k_1 x_1 + k_2 x_2 )(x_1 + x_2) + 2 (x_1^2 + x_1 x_2 + x_2^2).
            \label{3.10}
     \end{eqnarray} 
 We note that relation (\ref{3.9}) is obeyed for  $ \alpha {\cal P} < 0$. 
 Let us prove that 
 \begin{equation}
         {\cal P} < 0.       \label{3.10a}
 \end{equation}
 Indeed, using relation (\ref{3.15}), or  $m + k_1 x_1 + k_2 x_2 = 1 +  x_1 +  x_2$,
 we get 
   \begin{eqnarray}
       {\cal P}  =   (1 +  x_1 +  x_2 )^2  - (m + k_1 x_1^2 + k_2 x_2^2) 
       \nonumber \\
       - 2 (1 +  x_1 + x_2 )(x_1 + x_2) + 2 (x_1^2 + x_1 x_2 + x_2^2) 
       \nonumber \\
       = 1 - m + (1 - k_1) x_1^2 +  (1 - k_2) x_2^2 < 0, 
                      \label{3.10b}
       \end{eqnarray} 
 for $m > 1$, $k_1 > 1$, $k_2 > 1$. 
 
  Hence, the solutions under consideration
 take place only if 
  \begin{equation}
       \alpha   >  0.       \label{3.10c}
  \end{equation}
    
    The calculations gives us the following relation for the vector 
    $v$ from (\ref{3.1}) 
               \begin{equation}
             G_{ij} v^i v^j = m H^2 + k_1 h_1^2 + k_2 h_2^2 - (m H + k_1 h_1 + k_2 h_2)^2                
             \label{3.6a} 
             \end{equation}                                                        
               and
                \begin{eqnarray}
                 G_{ijkl} v^i v^j v^k v^l =    m (m-1) (m-2) (m - 3) H^4      \nonumber \\         
                + 4 m (m-1) (m-2)  H^3 (k_1 h_1 + k_2 h_2)                   \nonumber \\   
    + 6 m (m-1) H^2 [k_1 (k_1 -1) h^2_1  + 2 k_1 k_2 h_1 h_2 + k_2 (k_2 - 1) h^2_2 ]
                                                                              \nonumber \\
                + 4 m H [ k_1 (k_1 -1)(k_1 -2) h^3_1   
                   + 3 k_1 (k_1 -1)k_2  h^2_1 h_2 
                                                                              \nonumber \\
                   +  3 k_1 k_2 (k_2 -1) h_1 h_2^2 
                   +  k_2 (k_2 -1)(k_2 - 2) h^3_2]  
                                                                              \nonumber \\   
                     + k_1 (k_1 -1)(k_1 -2) (k_1 - 3) h^4_1 
                    + 4 k_1 (k_1 -1)(k_1 -2) k_2  h^3_1 h_2
                                                                              \nonumber \\
                    + 6 k_1 (k_1 -1)k_2 (k_2 - 1)  h^2_1 h_2^2
                    + 4 k_1 k_2 (k_2 -1)(k_2 -2)  h_1 h_2^3 
                                                                              \nonumber \\
                    + k_2 (k_2 -1)(k_2 -2) (k_2 - 3) h^4_2 .                   \label{3.6b}                         
                \end{eqnarray}
                  This may be obtained by using  the relation from Ref. \cite{Iv-10}
                    \begin{equation}
                   G_{ijkl}v^i v^j v^k v^l =
                   S_1^4 -  6  S_1^2 S_2 + 3 S_2^2 + 8  S_1 S_3 - 6 S_4.
                    \label{3.7}
                  \end{equation}

  Due to  (\ref{2.3}),   (\ref{3.6a}) and (\ref{3.6b}),  the equation (\ref{3.4E}) reads 
   \begin{eqnarray}
   2 \Lambda = - G_{ij} v^i v^j +
     \alpha   G_{ijkl} v^i v^j v^k v^l 
                             \nonumber \\
     = H^2 V_1  +  \alpha H^4 V_2, 
                         \label{3.11}    
   \end{eqnarray}  
 where 
  \begin{eqnarray}
   V_1 = V_1(x_1,x_2) = V_1(x_1,x_2, m, k_1, k_2) 
    \nonumber \\
   =  - m - k_1 x_1^2 - k_2 x_2^2 + (m  + k_1 x_1 + k_2 x_2)^2 \label{3.12a} 
   \end{eqnarray}
   and 
   \begin{eqnarray}                          
   V_2 = V_2(x_1,x_2) = V_2(x_1,x_2, m, k_1, k_2) 
       \nonumber \\          
   = [m]_4   + 4 [m]_3  (k_1 x_1 + k_2 x_2)                   
       + 6 [m]_2 \left( [k_1]_2 x^2_1  + 2 k_1 k_2 x_1 x_2 + [k_2]_2  x^2_2 \right)
                                                               \nonumber \\
      + 4 m \left( [k_1]_3 x^3_1  +  3 [k_1]_2 k_2  x^2_1 x_2 
           +  3 k_1 [k_2]_2  x_1 x_2^2 +  [k_2]_3 x^3_2 \right)  
                                                                \nonumber \\   
       + [k_1]_4 x^4_1 +  4 [k_1]_3 k_2  x^3_1 x_2                                                                 
             + 6 [k_1]_2 [k_2]_2  x^2_1 x_2^2
             + 4 k_1 [k_2]_3  x_1 x_2^3 + [k_2]_4 x^4_2.       \label{3.12b}                              
                                                         \end{eqnarray} 
Here we use the notation $[N]_k = N (N-1)... (N - k +1)$.

Using (\ref{3.9}) we get 
\begin{equation}
  \lambda = \alpha \Lambda 
     = - \frac{V_1}{4 {\cal P} }  +  \frac{V_2}{8 {\cal P}^2 }, 
                         \label{3.13}    
   \end{equation} 
or, equivalently, 
\begin{equation}
  V_2(x_1,x_2) - 2 {\cal P}(x_1,x_2) V_1 (x_1,x_2)  -   8 ({\cal P}(x_1,x_2))^2 \lambda = 0. 
                         \label{3.14}    
 \end{equation} 
Thus, we are led to polynomial equation in variables $x_1, x_2$ of fourth order or less
(depending upon  $\lambda$).

We call  relations  (\ref{3.15}), (\ref{3.14}),    as a master equations. The set of these
equations may solved in radicals. Indeed, solving eq. (\ref{3.15}) 
\begin{equation}
 x_2 =  x_2(x_1) = - \frac{m -1}{k_2 -1}  -  \frac{k_1 - 1}{k_2 -1} x_1  \label{3.16}
\end{equation}
and substituting into eq. (\ref{3.14})  we obtain another (master) equation in $x_1$
 \begin{equation}
   V_2(x_1,x_2(x_1)) - 2 {\cal P}(x_1,x_2(x_1)) V_1 (x_1,x_2(x_1)) 
    -   8 ({\cal P}(x_1,x_2(x_1)))^2 \lambda = 0, 
                          \label{3.17}    
  \end{equation}  
which is of fourth order or less depending upon the value of $\lambda$. It may solved in radicals
for all $m > 1$, $k_1 > 1$ and $k_2 > 1$. Here we do not try to write the explicit solution 
for general setup. It seems more effective for any given dimensions $m $, $k_1$ and $k_2 $ 
to find the solutions just by using Maple or Mathematica. An example of solution with $k_1 = k_2$ will 
be considered below.    

In what follows we  use the  identity
 \begin{eqnarray}
 - (k_2 - 1) {\cal P}(x_1,x_2(x_1)) = (k_1 - 1)(k_1 + k_2 -2)x_1^2  \nonumber \\ 
   + 2(m-1)(k_1 -1)x_1 + (m-1)(m + k_2 - 2),
 \label{3.18} 
\end{eqnarray}
following from  (\ref{3.10b}) and (\ref{3.16}).

\section{The case $k_1  \neq k_2$}

Here we put the following restriction $k_1  \neq k_2$.  
We write  relation (\ref{3.13}) as
\begin{equation}
  \lambda = f(x_1) \equiv   
      - \frac{V_1 (x_1,x_2(x_1))}{4 {\cal P}(x_1,x_2(x_1))} 
      +  \frac{V_2(x_1,x_2(x_1))}{8 ({\cal P}(x_1,x_2(x_1)))^2 }. 
                         \label{3.13a}    
   \end{equation} 

Using relation (\ref{3.16}) we rewrite the restrictions (\ref{3.3a}) (respectively) as follows
 \begin{equation}
        x_1 \neq X_1, \quad  x_1 \neq X_2, \quad  x_1 \neq X_3, \quad  x_1 \neq X_4,
      \label{3.3b}
    \end{equation}
where
 \begin{eqnarray}
 X_1 = 1,
   \label{3.x1} \\
 X_2 = -\frac{m + k_2 -2}{k_1 -1},
   \label{3.x2} \\
 X_3 = -\frac{m-1}{k_1 + k_2 -2},
   \label{3.x3} \\
 X_4 = \frac{m - k_2}{k_2 - k_1}.
   \label{3.x4}
 \end{eqnarray}

\subsection{ Extremum points }

The calculations  give us  

\begin{equation}
\frac{df}{dx_1}= \frac{C(m, k_1, k_2)(x_1 - X_1)(x_1 - X_2)(x_1 - X_3)(x_1 - X_4)}{\bigg(
  - (k_2 - 1) {\cal P}(x_1,x_2(x_1)) \bigg)^3},
  \label{3.f}
 \end{equation}
 where
 \begin{equation}
 C(m, k_1, k_2) = (m-1)(k_1 - 1)^2(k_2 - k_1)(k_1 + k_2 -2)
  \label{3.c}
 \end{equation}
  and $X_1, X_2,  X_3, X_4$ are defined in (\ref{3.x1})-(\ref{3.x4}). Thus, the 
  points of extremum of the function  $f(x_1)$ are excluded from our consideration 
  due to restrictions  (\ref{3.3}).
 
For the values $\lambda_i = f(X_i)$, $i =1,2,3,4$, we get 

\begin{eqnarray}
\lambda_1 = \lambda_1 (m,k_1,k_2) = \frac{u(k_2,m + k_1)}{8(m + k_1 + k_2 - 3)(m + k_1 -2)(k_2 -1)},
   \label{3.l1L} \\
\lambda_2 = \lambda_2 (m,k_1,k_2) = \frac{ u(k_1, m + k_2)}{8(m + k_1 + k_2 - 3)(m + k_2 -2)(k_1 -1)},
   \label{3.l2L} \\
\lambda_3 = \lambda_3 (m,k_1,k_2) = \frac{u(m, k_1 + k_2)}
{8(m-1) (k_1 + k_2 -2)(m + k_1 + k_2 - 3)},
   \label{3.13L} \\
 \lambda_4 = \lambda_4 (m,k_1,k_2) = \frac{v(m,k_1,k_2)}{8 w(m,k_1,k_2)},
   \label{3.l4L}
 \end{eqnarray}
where
\begin{eqnarray}
u(m, l) = l m^2 + (l^2 - 8l +8)m + l(l -1), \label{3.u} \\
v(m,l,k) = (k+ l)m^2 + (m + l )k^2 +(m + k)l^2 - 6mlk, \label{3.v} \\
w(m,l,k) = (k+ l - 2)m^2 + (m + l - 2)k^2 +(m + k - 2)l^2
\nonumber \\
   + 2 m l + 2mk + 2lk   -  6mlk. \label{3.w}
\end{eqnarray} 

We note that 
 \begin{equation}
   \lambda_i = \lambda_i(m,k_1,k_2)> 0 \label{3.L} 
 \end{equation}
for all $m > 1$, $k_1 > 1$, $k_2 > 1$, $i = 1,2,3,4$. 

For $i = 1,2,3$ this relation follows from the 
 \begin{equation}
 u(m, l) > 0  \label{3.uu}
 \end{equation}
for  $m > 1$ and  $l > 1$. Indeed,  for $m \ge 4$, $l \ge 4$  we get 
$u(m,l) = ml(m + l- 8) + 8m +l^2-l > 0$ and $u(4,3) = 26$, $u(3,4) = 24$, $u(3,3) = 12$,
$u(3,2) = 8$, $u(2,3) = 4$, $u(2,2) = 2$.
For $i =4$ the relation (\ref{3.L}) follows from
the inequalities  
\begin{eqnarray}
v(m,l,k) > 0, \label{3.vv} \\
w(m,l,k) > 0, \label{3.ww}
\end{eqnarray}
which are valid for natural numbers $m,l,k$ obeying: $m>1$, $l>1$, $k > 1$ and 
either $m \neq l$, or  $m \neq k$, or $l \neq k$. This is proved in Appendix.  

We also note that the following symmetry identities take place for the functions $\lambda_i(m,k_1,k_2)$, 
$i = 1,2,3$,    
\begin{eqnarray}
\lambda_1 (m,k_1,k_2) = \lambda_2 (m,k_2,k_1) = \lambda_3 (k_2,m,k_1),
   \label{3.123L} \\
 \lambda_3 (m,k_1,k_2) = \lambda_3 (m,k_2,k_1).
   \label{3.33L}
    \end{eqnarray}

The function $\lambda_4 (m,k_1,k_2)$ is symmetric with respect to variables since the functions 
$v (m,k_1,k_2)$ and $w (m,k_1,k_2)$ are symmetric.

For $x_1 \to \pm \infty $ we get
\begin{equation}
  \lambda_{\infty}= \lim_{x_1 \to  \infty } f(x_1) =   
  \frac{(k_1 + k_2 -6)k_1k_2 + k_1^2 + k_2^2 + k_1 + k_2}{8(k_1-1)(k_2-1)(k_1 + k_2 -2)}.
  \label{3.in}
\end{equation}

 It may be readily verified that 
\begin{equation}
  \lambda_{\infty} = \lambda_{\infty}(k_1, k_2) = \lambda_{\infty}(k_2, k_1) > 0,
  \label{3.inn}
\end{equation}
for all $k_1 > 1$ and $k_2 > 1$. Indeed, 
$(k_1 + k_2 -6)k_1k_2 + k_1^2 + k_2^2 + k_1 + k_2 = (k_1 + k_2 -4)k_1k_2 + (k_1 - k_2)^2  + k_1 + k_2 > 0$
for $k_1 \geq 2$ and $k_2 \geq 2$.

The points of extremum obey the following relations

\begin{eqnarray}
X_2 - X_1 = - \frac{m + k_1 +  k_2 -3}{k_1 -1},
   \label{3.X21} \\
X_3 - X_1 = -\frac{m+ k_1+ k_2 -3}{k_1 + k_2 -2},
   \label{3.X31} \\
X_3 - X_2 = \frac{(m + k_1 +  k_2 -3)(k_2 -1)}{(k_1 +  k_2 -2)(k_1 -1)},
   \label{3.X32} \\ 
 X_4 - X_1 = - \frac{m+ k_1 - 2k_2}{k_1 - k_2}, 
   \label{3.X41}  \\
X_4 - X_2 = \frac{(m - 2k_1 + k_2)(k_2 -1)}{(k_1 -1)(k_2 - k_1)},
   \label{3.X42} \\
X_4 - X_3 = \frac{(2m-k_1-k_2)(k_2-1)}{(k_1+k_2-2)(k_2 - k_1)}.
   \label{3.X43}
 \end{eqnarray}

It follows from definitions of $X_i$ and  (\ref{3.X21}), (\ref{3.X31}), (\ref{3.X32}) that
 \begin{equation}
       X_2 <  X_3 < 0 < X_1 = 1
   \label{3.X123}
 \end{equation}
for all $m > 1$, $k_1 > 1$ and $k_2 > 1$.

The corresponding relations for $\lambda_i - \lambda_j$ 
have the following form 
\begin{eqnarray}
\lambda_2 - \lambda_1 = 
\frac{(m- 1)(k_2 - k_1)(m+ k_1 + k_2 -3)}{4 (k_1 -1)(k_2 - 1)(m+k_1 -2)(m+ k_2 -2)}, 
   \label{3.L21} \\
\lambda_3 - \lambda_1 = 
\frac{(k_1 - 1)(k_2 - m)(m+ k_1 + k_2 -3)}{4 (m-1)(k_2 - 1)(m+ k_1 -2)(k_1 + k_2 -2)}, 
   \label{3.L31} \\
\lambda_3 - \lambda_2 = 
\frac{(k_2 - 1)(k_1 - m)(m+ k_1 + k_2 -3)}{4(m-1)(k_1 - 1)(m+ k_2 -2)(k_1 + k_2 -2)},
   \label{3.L32} \\
\lambda_4 - \lambda_1 =
\frac{(m-1)(k_1 - 1)(2k_2 - k_1 -m)^3}{4  (m + k_1 -2)(k_2 -1)(m + k_1 + k_2 -3) w},
   \label{3.L41} \\
 \lambda_4 - \lambda_2 = 
 \frac{(m-1)(k_2 - 1)(2k_1 - m - k_2)^3}{4  (m + k_2 -2)(k_1 -1)(m + k_1 + k_2 -3) w},
   \label{3.L42} \\
\lambda_4 - \lambda_3 = 
 \frac{(k_1-1)(k_2 - 1)(2m - k_1 - k_2)^3}{4 (k_1 + k_2 -2)(m -1)(m + k_1 + k_2 -3) w},
   \label{3.L43}
 \end{eqnarray}
where $w = w(m, k_1, k_2)$ is defined in (\ref{3.w}).

Here and in what follows up to the Section 4 we put  that
 \begin{equation}
   1 < m < k_1 < k_2.
  \label{3.mk1k2}
 \end{equation}

Using (\ref{3.L21}), (\ref{3.L32})  and (\ref{3.mk1k2}) we get
\begin{equation}
   0 < \lambda_1 < \lambda_2 < \lambda_3.
  \label{3.Lmk1k2}
 \end{equation}

Analogously, using (\ref{3.L41}), (\ref{3.L43})  and (\ref{3.mk1k2}) we get
\begin{equation}
   0 < \lambda_1 < \lambda_4 < \lambda_3.
  \label{3.Lm4k2}
 \end{equation}

It follows from (\ref{3.X42}), (\ref{3.L42}) and (\ref{3.mk1k2}) that 
 \begin{equation}
  (A_{+}) \quad X_4 < X_2, \qquad  \lambda_4 > \lambda_2, 
  \quad {\rm for} \quad 2k_1 - m - k_2 > 0,
  \label{3.XL1}
 \end{equation}

\begin{equation}
  (A_{-}) \quad X_4 > X_2, \qquad  \lambda_4 < \lambda_2, 
  \quad {\rm for} \quad 2k_1 - m - k_2 < 0,
  \label{3.XL2}
 \end{equation}
and 
\begin{equation}
  (A_{0}) \quad X_4 = X_2, \qquad  \lambda_4 = \lambda_2, 
  \quad {\rm for} \quad 2k_1 - m - k_2 = 0.
  \label{3.XL3}
 \end{equation}

The  graphical representations of the function $\lambda = f(x_1)$ 
for $(m, k_1, k_2) = (4, 6, 7), (4, 5, 7), (4, 5, 6)$ are  given at Figures 1, 2 and 3,
respectively. These three sets obey the inequalities (\ref{3.XL1}),
(\ref{3.XL2}) and (\ref{3.XL3}), respectively.

 \begin{figure}[!h]
	\begin{center}
		\includegraphics[width=0.75\linewidth]{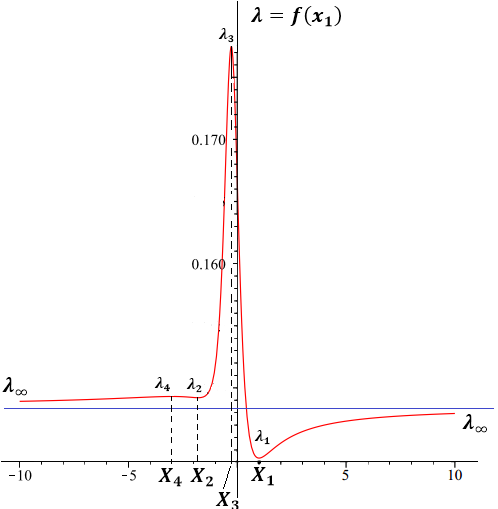}
		\caption{The function $\lambda = f(x_1)$ for  $m=4$,  $k_1=6$, $k_2 = 7$.}
		\label{rfig:1}
	\end{center}
 \end{figure}

 \begin{figure}[!h]
	\begin{center}
		\includegraphics[width=0.75\linewidth]{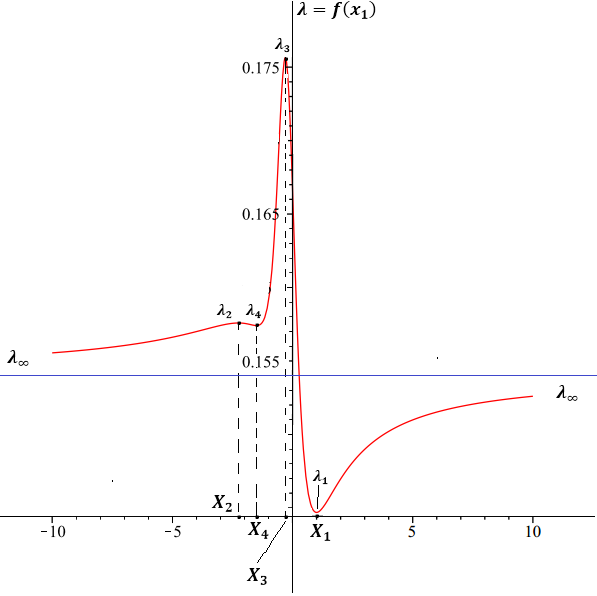}
		\caption{The function $\lambda = f(x_1)$ for  $m=4$,  $k_1=5$, $k_2 = 7$.}
		\label{rfig:2}
	\end{center}
 \end{figure}
 
 \begin{figure}[!h]
 	\begin{center}
 		\includegraphics[width=0.75\linewidth]{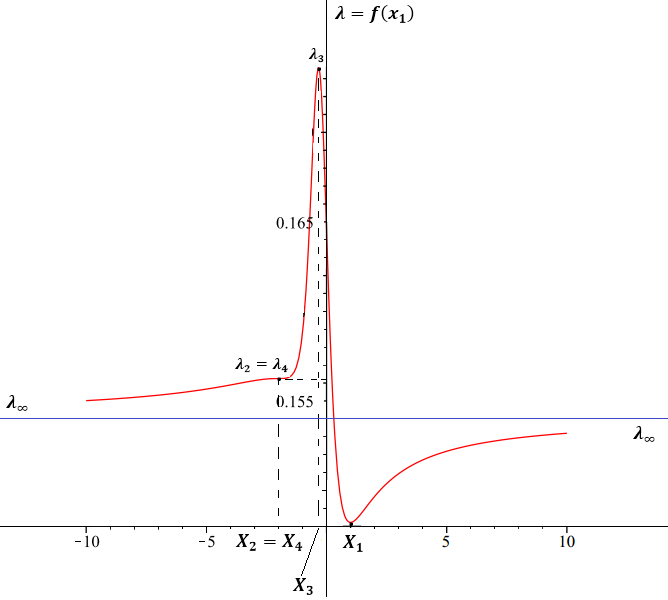}
 		\caption{The function $\lambda = f(x_1)$ for  $m=4$,  $k_1=5$, $k_2 = 6$.}
 		\label{rfig:3}
 	\end{center}
  \end{figure}

For $\lambda_{i} - \lambda_{\infty}$ we obtain
\begin{eqnarray}
\lambda_1 - \lambda_{\infty} = \frac{(m - 1)z_1}{4(k_1 -1)(k_1 + k_2 - 2)(m+ k_1 -2)(m + k_1 + k_2 - 3)}, \
   \label{3.L1} \\
 \lambda_2 - \lambda_{\infty} = 
 \frac{(m - 1)z_2}{4 (k_2 -1)(k_1 + k_2 - 2)(m+ k_2 -2)(m + k_1 + k_2 - 3)}, \
   \label{3.L2} \\
\lambda_3 - \lambda_{\infty} =
 \frac{z_3}{4 (m-1)(k_1 - 1)( k_2 - 1)(k_1 + k_2 -2)(m + k_1 + k_2 -3)}, \
   \label{3.L3} \\
 \lambda_4 - \lambda_{\infty} =
   \frac{(m- 1)z_4}{4  (k_1 - 1)(k_2 - 1)(k_1 + k_2 -2) w}, \quad
   \label{3.L4}
 \end{eqnarray}
where
\begin{eqnarray}
 z_1 = (2k_1-k_2 - 1)m - 2k_1k_2 -  4k_1  + 5k_2 -  k_2^2  + 2k_1^2,
  \label{3.z1} \\   
 z_2 = (2k_2 - k_1 -1)m - 2k_1k_2 - 4k_2 + 5k_1 - k_1^2 + 2k_2^2,
  \label{3.z2} \\  
  z_3 = - (k_1 - k_2)^2 m^2 - ((k_1^2 - 6k_1 +6)k_1 + (k_2^2 - 6k_2 +6)k_2 \nonumber \\
  + (k_1 + k_2)k_1k_2 - 4)m -  2(k_1 + k_2) \nonumber \\
  +  (12 - 6k_1 - 6k_2  + (k_1 + k_2)^2 )k_1k_2,
  \label{3.z3} \\
   z_4 = - (k_2 - k_1)^2((k_1 + k_2 -2)m + (k_1 - k_2)^2 + k_1 - 2k_1k_2 + k_2 ).
 \label{3.z4} 
 \end{eqnarray}

It follows from  (\ref{3.L1}), (\ref{3.L3}) and  inequalities $z_1 < 0$, 
  $z_3 > 0$, proved in Appendix,  that
\begin{equation}
 \lambda_1 < \lambda_{\infty} < \lambda_3.
 \label{3.L1inf3}
\end{equation}

For our restriction  (\ref{3.mk1k2}) we obtain from (\ref{3.c})
\begin{equation}
 C(m, k_1, k_2) >0. 
  \label{3.cc}
 \end{equation}
 
 In what follows we use the relation (\ref{3.f}) and inequalities 
 (\ref{3.10a}) and  (\ref{3.cc}).
  
  We find that (in all cases) the function $\lambda = f(x_1)$
 is monotonically increasing in the interval $(X_1 = 1, + \infty)$ from 
 $\lambda_1$ to  $\lambda_{\infty}$ and it is monotonically decreasing 
 in the interval $(X_3, X_1)$ from  $\lambda_3$ to  $\lambda_{1}$.
 
 In the case $(A_{+})$ the function $\lambda = f(x_1)$ 
 is monotonically increasing in the intervals $(-\infty, X_4)$ and 
 $(X_2, X_3)$ from $\lambda_{\infty}$ to  $\lambda_4$ and  
 from $\lambda_{2}$ to  $\lambda_3$, respectively, while 
 it is monotonically decreasing  in the interval 
 $(X_4, X_2)$ from  $\lambda_4$ to  $\lambda_{2}$ (see Figure 1).
 In this case the points $X_1$ and $X_2$ are points of local minimum and points 
 $X_3$ and $X_4$ are points of local maximum. 
 
 For the case $(A_{-})$ the function $\lambda = f(x_1)$ 
 is monotonically increasing in the intervals $(-\infty, X_2)$ and 
 $(X_4, X_3)$ from $\lambda_{\infty}$ to  $\lambda_2$ and  
 from $\lambda_{4}$ to  $\lambda_3$, respectively, while  
 it is monotonically decreasing  in the interval 
 $(X_2, X_4)$ from  $\lambda_2$ to  $\lambda_{4}$ (see Figure 2).
 The points $X_1$ and $X_4$ are points of local minimum and points 
  $X_2$ and $X_3$ are points of local maximum. 
  In this case $\lambda_2 > \lambda_{\infty}$.
 
 In the case $(A_{0})$ the function $\lambda = f(x_1)$ 
  is monotonically increasing in the intervals $(-\infty, X_3)$
   from $\lambda_{\infty}$ to  $\lambda_3$, respectively  (see Figure 3).  
   For this case the point $X_1$ is the  point of local minimum,  the point 
    $X_3$ is a point of local maximum and the point $X_2 = X_4$ is a point
    of inflection.
 
  Using the inequalities (\ref{3.Lmk1k2}), (\ref{3.Lm4k2})  and  (\ref{3.L1inf3}) we get
  from the behaviour of the function $f(x_1)$ mentioned above that 
 $X_3$ is the point of absolute maximum and $X_1$ is the point of absolute minimum,
 i.e.    
 \begin{equation}
       \lambda_1   \leq   \lambda = f(x_1) \leq \lambda_3
   \label{3.Lminmax}
  \end{equation}
 for all $x_1 \in \R$. 
 Due to (\ref{3.3b}) the points  $X_1, X_2, X_3, X_4$ are forbidden for our
 consideration. We get 
 \begin{equation}
 \lambda_1   < \lambda = f(x_1) < \lambda_3 
  \label{3.Lminmaxno}
 \end{equation}
 for all $x_1 \neq X_1, X_2, X_3, X_4$. 
 Let us denote the set of definition of the fuction $f$
 for our consideration
 $(-\infty, \infty)_{*} \equiv \{x| x \in \R, x \neq X_1, X_2, X_3, X_4 \}$.
 Since the function $f(x_1)$ 
 is continuous one the image of the function $f$ 
 (due to intermediate value theorem) is 
  \begin{equation}
 f((-\infty, \infty)_{*}) = (\lambda_1,  \lambda_3). 
  \label{3.ffa}
  \end{equation}
 Thus, we a led the following proposition.
 
 {\bf Proposition 1.} {\em  The solutions to equations
 (\ref{2.3}), (\ref{2.4})  for  ansatz  (\ref{3.1}) with   $1 < m < k_1 < k_2$ 
 obeying the inequalities $H \neq 0$,  $H \neq h_1$, $H \neq h_2$, $h_1 \neq h_2$ and 
 $S_1 = m H + k_1 h_1 + k_2 h_2 \neq 0$ do exist if and only if $\alpha > 0$ and 
 \begin{equation}
    0 < \lambda_1 <  \alpha \Lambda < \lambda_3,
   \label{3.L13}
  \end{equation}
   where $\lambda_1$ and $\lambda_3$ 
  are defined in (\ref{3.l1L}) and (\ref{3.13L}), respectively. 
  In this case  $x_1 = h_1/H \neq X_1, X_2, X_3, X_4$ 
 (see (\ref{3.x1}), (\ref{3.x2}), (\ref{3.x3}), (\ref{3.x4})),  
  $x_2 = h_2/H = x_2(x_1)$ is given by  (\ref{3.16}), $x_1$ obeys the 
  polynomial master equation (\ref{3.17}) (of fourth order or less) and 
  $H^2$ is given by (\ref{3.9}) and (\ref{3.10}). } 
 
 {\bf The case $H = 0$.} 
 It may verified  that in the case $H = 0$ the solutions under consideration  take place only if $\alpha > 0$,  and 
   \begin{equation}
    \alpha \Lambda = \lambda_{\infty}(k_1,k_2) 
    = \frac{(k_1 + k_2 -6)k_1k_2 + k_1^2 + k_2^2 + k_1 + k_2}{8(k_1-1)(k_2-1)(k_1 + k_2 -2)} > 0, 
   \label{3.R1.0}
   \end{equation}
 where $k_1 \neq k_2$. Indeed (\ref{3.4L}) is equivalent to   
 $(k_1 - 1) h_1 + (k_2 - 1) h_2 = 0$,                                        
 while (\ref{3.4Q})  reads as 
  $(k_1 - 1) (h_1)^2 + (k_2 - 1) (h_2)^2 = 1/(2 \alpha)$.
 These relations  imply  $\alpha > 0$ and
 \begin{eqnarray}
 h_1 = \pm \left(\frac{k_2 -1}{2 \alpha  (k_1 -1) (k_1 + k_2 - 2)}\right)^{1/2},
    \label{3.R1.1} \\ 
 h_2 = \mp \left(\frac{k_1 -1}{2 \alpha  (k_2 -1) (k_1 + k_2 - 2)}\right)^{1/2}
   \label{3.R1.2}.
 \end{eqnarray}
 The substitution of these values of $h_1$ and $h_2$, and $H=0$ into equation (\ref{3.4E})
 gives us (due to  (\ref{3.6a}) and (\ref{3.6b}))  relation  (\ref{3.R1.0}).  
 
\section{The case $k_1 = k_2$}

Here we consider the case $m > 1$, $k_1 = k_2 = k > 1$ and $H \neq 0$.
We get from (\ref{3.15}) 
\begin{equation}
    m -1  + (k - 1)(x_1 +  x_2) = 0.  \label{4.1}
\end{equation}
In this case relation (\ref{3.10b}) implies
\begin{equation}
 {\cal P}  = 1 - m + (1 - k) (x_1^2 +  x_2^2). 
 \label{4.2}
\end{equation} 

The solutions under consideration take place for 
\begin{equation}
    m \neq k  \label{4.1m}
\end{equation}
and $\alpha > 0$ (see Section 2).

Let us denote 
\begin{equation}
   X \equiv \alpha H^2,    \label{4.3}
\end{equation}
$\alpha > 0$.
It follows from (\ref{3.9})
\begin{equation}
   X  {\cal P} = - \frac{1}{2}.    \label{4.4}
\end{equation}
Due to (\ref{4.3}) we have 
  \begin{equation}
     H = \varepsilon_0 \sqrt{X/\alpha}, \qquad \varepsilon_0 = \pm 1 . 
        \label{4.5}
  \end{equation}

The substitution of relations (\ref{4.1}), (\ref{4.2})
into formulae  (\ref{3.12a}), (\ref{3.12b}) gives us
\begin{eqnarray}
V_1 = [(m-1)(m-k) + {\cal P} k (k-1)]/(k-1)^2,     \label{4.6} \\  
V_2  = [- (m-1)(m-k)(m+k-2)(m+2k-3)  \nonumber \\
         + 3 {\cal P}^2 (k-1)^2 k  ]/(k - 1)^3.       \label{4.7}
\end{eqnarray}
  Using (\ref{4.4}) we rewrite relation (\ref{3.13}) as
  \begin{equation}
  2 \lambda = 2 \alpha \Lambda  =  X V_1 +  X^2 V_2. 
                              \label{4.8}    
  \end{equation} 
 This relation may be written as quadratic relation  
 \begin{equation}
  A X^2 + B X + C = 0,          \label{4.9}
  \end{equation} 
  where
  \begin{eqnarray}
 A = (m-1)(m-k)(m+k-2)(m+2k-3), \label{4.9A} \\
 B = -(m-1)(m-k)(k-1),   \qquad      \label{4.9B} \\
 C=  - \frac{1}{4} k(k-1)^2 + 2 \lambda (k - 1)^3. \quad  \label{4.9C} \\  
 \end{eqnarray}
Due to (\ref{4.1m})  $A \neq 0$. 
The discriminant $D = B^2 - 4 A C$ has the folowing form
 \begin{equation}
   D =  (m-1)(m-k)(k-1)^2 (F   - 8 \lambda f),   \label{4.10}
   \end{equation} 
  where
 \begin{eqnarray}
   F = F(m,k) = (m-1)(m-k) + (m+ k-2)(m+2k-3)k, \label{4.11}      \\
   f = f(m,k) = (m+k-2)(m+2k-3)(k - 1) > 0  \label{4.12} . 
  \end{eqnarray}
 
{\bf Lemma}. 
{\it $F = F(m,k) >0$ for all $m > 1$, $k > 1$ and $k \neq m$.} 

{\bf Proof.}
For $m > k$ we have a sum of two positive terms in 
(\ref{4.11}) and hence $F > 0$ in this case.
For $k > m$,  we denote  $k = m + p$, $p > 0$.  
We obtain 
 \begin{eqnarray}
F = (m-1)(-p) + (2m+p-2)(3m + 2 p - 3)(m + p) =
 \nonumber        \\
= (m-1)(-p) + (2(m -1)+p)(3(m -1) + 2 p)(m + p) = 
\nonumber        \\
= 2 p^3+(9m-7)p^2+ (m-1)(13m-7) p + 6m(m -1)^2.
\label{4.13}
 \end{eqnarray}
 Due to $m > 1$ and $p > 0$ we have a sum of three
 positive terms in (\ref{4.13}) and hence $F > 0$ for $k > m$.

The solution  to eq. (\ref{4.9}) reads 
\begin{equation}
  X = (- B + \bar{\varepsilon}_1 \sqrt{D})/(2A), \qquad \bar{\varepsilon}_1 = \pm 1. 
        \label{4.14}
 \end{equation}

We are seeking real soutions which obey two restrictions  
\begin{eqnarray}
 D  > 0, \label{4.15D} \\
 X > 0.   \label{4.15X} 
 \end{eqnarray}
 Here the case $D = 0$ is excluded from the consideration since as it will be shown later
 it implies either $x_1 = 1$ or $x_2 = 1$, which contradict  restrictions (\ref{3.3a}).

 The inequality (\ref{4.15D}) may be rewritten as 
 \begin{eqnarray}
  \lambda  < \lambda_1 \ {\rm for} \ m > k,   \label{4.16a} \\
  \lambda  > \lambda_1 \ {\rm for} \ m < k,   \label{4.16b}   
  \end{eqnarray}
 where 
 \begin{equation}
   \lambda_1 = \lambda_1(m,k,k) = F(m,k)/(8 f(m,k)). 
         \label{4.17}
  \end{equation}
For definition of $\lambda_1(m,k,l)$ see (\ref{3.l1L}).

The set of two equations (\ref{4.1}) and (\ref{4.2}) 
have the following solutions
\begin{eqnarray}
x_1 = -(\varepsilon_2 \sqrt{E}   
      +m-1)/(2k-2), \label{4.18a} \\
x_2 = -(- \varepsilon_2 \sqrt{E} + m-1)/(2k-2), 
\label{4.18b}
\end{eqnarray}
where $\varepsilon_2 = \pm 1$ and 
\begin{eqnarray}
  E = -(m-1)(m+2k-3) - 2 {\cal P} (k - 1)  \nonumber \\
    =  (k-1) X^{-1} - (m-1)(m+2k - 3). 
  \label{4.19}
\end{eqnarray}
Here we put  
 \begin{equation}
   E > 0  \label{4.20}
  \end{equation}
since $E = 0$ implies the identity $x_1 = x_2$ 
which is excluded by restrictions (\ref{3.3a}). The relations 
(\ref{4.15X}) and (\ref{4.20}) may be written as 
\begin{equation}
   0 < X < \frac{k-1}{(m-1)(m+2k - 3)}.  \label{4.21}
  \end{equation}

Now we explain why the case $D = 0$ was excluded from our consideration.
Let us put  $D = 0$. Then we get from (\ref{4.14})
  \begin{equation}
  X = (- B)/(2A) = (k-1)/(2 (m + k -2)(m+2k - 3))  \label{4.22}
   \end{equation}
  and hence    
  \begin{equation}
  E = (m + 2k - 3)^2,  \label{4.23}
   \end{equation}
    which implies either $x_2 = 1$ for 
  $\varepsilon_2 = 1$ or  $x_1 = 1$ for $\varepsilon_2 = -1$. 
  But this is forbiden by first two inequalities in (\ref{3.3a}).
  
  Moreover, it is not difficult to verify that relations 
  (\ref{4.18a}), (\ref{4.18b}) and  (\ref{4.21}) imply
  all four inequalities in (\ref{3.3a}). Indeed, the violation
  of first two inequalities in (\ref{3.3a}) lead us either to 
  $x_1 = 1$ or  $x_2 = 1$ which may be valid only for $E$ 
  from  (\ref{4.23}) and  $\varepsilon_2 = - 1$ or $\varepsilon_2 =  1$,
  respectively. But due to definition (\ref{4.19}), relation
  (\ref{4.23}) implies (\ref{4.22}) and hence $D = 0$, 
  which contradict to relations   (\ref{4.18a}), (\ref{4.18b}). 
  The violation of the third inequality gives us $x_1 = x_2$ 
  which imply $E = 0$, but this is forbidden by (\ref{4.21}).
  Now, let us verify the last inequality in  (\ref{3.3a}).
  In our case it reads
  \begin{equation}
    x_1 +  x_2  \neq - \frac{m}{k}.    \label{4.24}
  \end{equation}
  From (\ref{4.18a}), (\ref{4.18b}) we obtain 
  \begin{equation}
      x_1 +  x_2  = - \frac{m - 1}{k - 1}.    \label{4.25}
    \end{equation}
   The relation is (\ref{4.24}) is satisfied due to (\ref{4.25}) and
   $m \neq k$.
   
   Now we  analyse the inequalities in (\ref{4.21}). We introduce 
   new  parameter 
   \begin{equation}
          \varepsilon_1 = \bar{\varepsilon}_1 {\rm sign} (m-k).  \label{4.26}
   \end{equation}
   Then relation (\ref{4.14}) reads as follows
 \begin{equation}
     X  = \frac{k-1}{2 (m + k -2)(m+2k - 3)} 
           + \varepsilon_1 \frac{\sqrt{D}}{2|A|},         \label{4.27}
 \end{equation}
 $\varepsilon_1 = \pm 1$. 

Let us consider the case  $\varepsilon_1 = - 1$. The second inequality in  (\ref{4.21})
      $X < \frac{k-1}{(m-1)(m+2k - 3)}$ is obeyed since $2 (m + k -2) > m - 1$. 
 Now we consider the first inequality $X > 0$. We get 
 \begin{equation}
     0 < \sqrt{D} < (m-1)|m-k|(k -1).         \label{4.28}
  \end{equation}
  Using the definition of $D$ in (\ref{4.10}) we obtain 
  \begin{equation}
    0 < (m-1)(m-k)(k -1)^2 (F - 8 \lambda f) <  (m-1)^2 |m-k|^2 (k - 1)^2.  \label{4.29}
  \end{equation} 
 
 Relations  (\ref{4.29}) read  as follows 
 \begin{eqnarray}
   F_{-} < 8 \lambda f <  F,  \ {\rm for} \  m > k,   \label{4.30a} \\
   F < 8 \lambda f <  F_{-}, \ {\rm for} \  m < k,   \label{4.30b} 
 \end{eqnarray} 
 where
 \begin{equation}
     F_{-} \equiv   F - (m-1)(m-k).  \label{4.30c}
  \end{equation}
 
 It may be verified  that 
   \begin{equation}
      \frac{F_{-}}{8f} = \frac{k}{8(k-1)} = \lambda_{\infty} = \lambda_{\infty}(k,k),
                         \label{4.30de}
   \end{equation}
 where $\lambda_{\infty}(k,l)$ is defined in  (\ref{3.in}).
 Using (\ref{4.17}) and (\ref{4.30de}) we rewrite relations (\ref{4.30a}), (\ref{4.30b})
 as follows
 \begin{eqnarray}
    \lambda_{\infty} <  \lambda  <   \lambda_{1},  \ {\rm for} \  m > k,   \label{4.30aa} \\
    \lambda_{1} <  \lambda  <  \lambda_{\infty}, \ {\rm for} \  m < k.   \label{4.30bb} 
  \end{eqnarray}

 Now, we put  $\varepsilon_1 = 1$. The inequality $X > 0$ is satisfied in 
 this case.  We should treat the inequality $X < \frac{k-1}{(m-1)(m+2k - 3)}$.
 We obtain 
 \begin{equation}
   0 < \sqrt{D} < |m-k|(m+2k-3)(k-1),     \label{4.31}
 \end{equation}
 or
 \begin{equation}
    0 < (m-1)(m-k) (F   - 8 \lambda f) <  |m-k|^2 (m+2k-3)^2.  
              \label{4.32}
   \end{equation}
 Relations  (\ref{4.32}) read  as follows 
  \begin{eqnarray}
   F_{+}  < 8 \lambda f <  F,  \ {\rm for} \  m > k,   \label{4.33a} \\
   F < 8 \lambda f <  F_{+} , \ {\rm for} \  m < k,   \label{4.33b} 
  \end{eqnarray}
where 
\begin{equation}
  F_{+} \equiv   F - (m-1)^{-1}(m-k)(m+2k-3)^2.          \label{4.33c}
 \end{equation}

 It may be verified  that 
   \begin{equation}
      \frac{F_{+}}{8f} = \lambda_{3} = \lambda_{3}(m,k,k),
                         \label{4.33de}
   \end{equation}
 where $\lambda_{3}(m,k,l)$ is defined in  (\ref{3.13L}).
 Using (\ref{4.17}) and (\ref{4.33de}) we rewrite relations 
 (\ref{4.33a}), (\ref{4.33b})  as follows
 \begin{eqnarray}
    \lambda_{3}  <  \lambda  <  \lambda_{1},  \ {\rm for} \  m > k,   \label{4.33aa} \\
    \lambda_{1} <  \lambda  <  \lambda_{3} , \ {\rm for} \  m < k .  \label{4.33bb} 
   \end{eqnarray}
 
 We note that  that 
 \begin{equation}
  \lambda_{1} <  \lambda_{\infty}  <  \lambda_{3}
                          \label{4.33dd}
  \end{equation}
 for $m < k$ (it proved in the previous section), while 
 \begin{equation}
   \lambda_{3} <  \lambda_{\infty}  <  \lambda_{1}
                           \label{4.33ee}
 \end{equation}
 for $k < m$. The inequalities in (\ref{4.33ee}) 
 follow from $F_{+} < F_{-} < F$ for $k < m$.

  {\bf Proposition 2.} {\em  The solutions to equations
  (\ref{2.3}), (\ref{2.4})  for  ansatz  (\ref{3.1}) with   $1 < m$, $1 < k_1 = k_2 = k$,
  $m \neq k$, 
  obeying the inequalities $H \neq 0$,  $H \neq h_1$, $H \neq h_2$, $h_1 \neq h_2$, 
  $S_1 = m H + k h_1 + k h_2 \neq 0$ do exist if and only if $\alpha > 0$, 
  \begin{equation}
     \lambda_1 < \lambda = \alpha \Lambda < \lambda_3
    \label{4.L13}
   \end{equation}
   for $m < k$  and 
   \begin{equation}
     \lambda_3 < \lambda = \alpha \Lambda < \lambda_1,
         \label{4.L31}  
    \end{equation}
   where $\lambda_1 = \lambda_1(k,k)$,  $\lambda_3=\lambda_3(k,k)$ 
   are defined in (\ref{3.l1L}) and  (\ref{3.13L}). 
   In this case $H$ obeys the relation (\ref{4.5}) with $X$ from 
      (\ref{4.27}),  $x_1 = h_1/H$ and 
   $x_2 = h_2/H$ are given by relations (\ref{4.18a}) and (\ref{4.18b}),
   $\lambda$ obeys  (\ref{4.30aa}),  (\ref{4.30bb})   for  $\varepsilon_1 = - 1$
   and (\ref{4.33aa}), (\ref{4.33bb}) for  $\varepsilon_1 =  1$
     with  $\lambda_{\infty} = \frac{k}{8(k-1)}$.  } 
 
 The restrictions on $\lambda$ for our solution may be explained just graphically
 as it was done in the previous section for $k_1  \neq  k_2$.  
 Indeed, for $k_1  =  k_2 = k \neq m$, $H \neq 0$  we have the same relation 
 (\ref{3.13a}) $\lambda = f(x_1)$,  where now
  \begin{equation}
  \frac{df}{dx_1}= \frac{\bar{C}(m, k) (x_1 - X_1)(x_1 - X_2)(x_1 - X_3)}{\bigg(
    - (k - 1) {\cal P}(x_1,x_2(x_1)) \bigg)^3}
    \label{3.ff}
   \end{equation}
 with 
  \begin{equation}
  \bar{C}(m, k) = 2 (m-1)(k - 1)^3 (k - m).
    \label{3.ccc}
   \end{equation}
 Here $x_2(x_1) = - \frac{m -1}{k -1}  -  x_1$
 and    restrictions (\ref{3.3a}) reads as follows
  \begin{equation}
         x_1 \neq X_1 =1, \quad  x_1 \neq X_2 = -\frac{m + k -2}{k -1},
          \quad  x_1 \neq X_3 = -\frac{m-1}{2 k -2},        \label{3.3bb}
     \end{equation}
 see (\ref{3.x1})-(\ref{3.x3}). 
 The fourth inequality in (\ref{3.3a}) is obeyed identically (it was checked above).  
 
 The points $X_1, X_2,  X_3$ are   points of extremum of the function $f(x_1)$.
 They are excluded from our consideration due to restrictions  (\ref{3.3bb}).
 The function $f(x_1)$ tends to $\lambda_{\infty}$ as $x_1$ tends to $\pm \infty$.
  
 Using  relations (\ref{3.ff}), (\ref{3.ccc}) and ${\cal P}(x_1,x_2(x_1)) < 0$  
 we get two cases. 
 
 For $1 < m < k$  the function  has two  points of minimum
 at $X_1$ and  $X_2$ with $\lambda_1 = f(X_1) = f(X_2) = \lambda_2 < \lambda_{\infty}$, and 
 the point of maximum at $X_3$ with $\lambda_3 = f(X_3)  > \lambda_{\infty}$. 
 See graphical representation of $f(x_1)$ for $m = 4$ and $k = 5$ at Figure 4. 

\begin{figure}[!h]
	\begin{center}
		\includegraphics[width=0.75\linewidth]{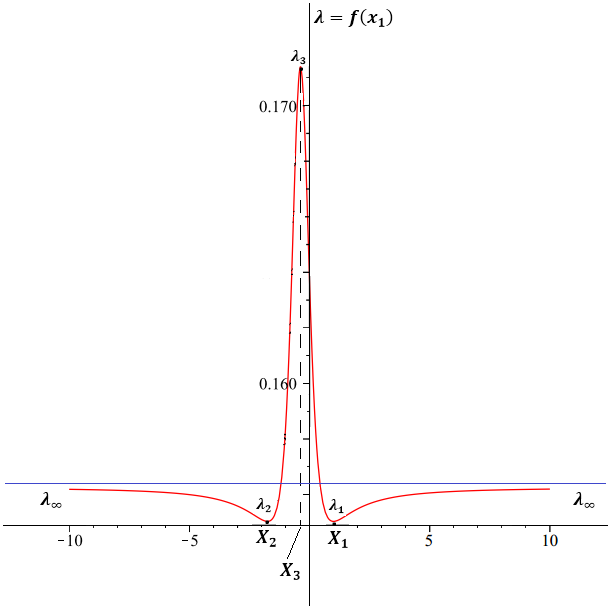}
		\caption{The function $\lambda = f(x_1)$ for  $m=4$,  $k_1= k_2 = 5$.}
		\label{rfig:4}
	\end{center}
 \end{figure}
    
  For $1 < k < m$  the function  has two  points of maximum
  at $X_1$ and  $X_2$ with $\lambda_1 = f(X_1) = f(X_2) = \lambda_2 > \lambda_{\infty}$, and 
  one point of minimum at $X_3$ with $\lambda_3 = f(X_3)  < \lambda_{\infty}$. 
  The graphical representation of $f(x_1)$ for $m = 5$ and $k = 4$ is depicted at Figure 5.

 We note that special solutions (e.g. stable ones) with $(m,k_1,k_2) = (3,4,4), (2,3,3), (4,3,3)$ were considered 
 earlier in  \cite{ErIv-19-GC}. 
 
 \begin{figure}[!h]
 	\begin{center}
 		\includegraphics[width=0.75\linewidth]{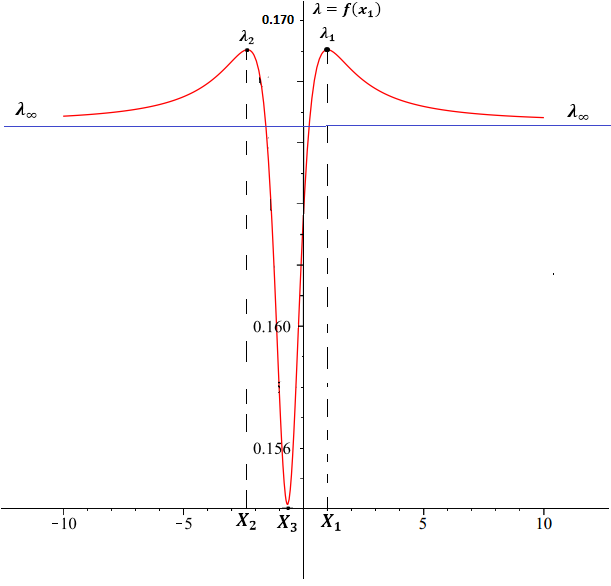}
 		\caption{The function $\lambda = f(x_1)$ for  $m=5$,  $k_1= k_2 = 4$.}
 		\label{rfig:5}
 	\end{center}
  \end{figure}

 {\bf The case $H = 0$.} For $k_1 = k_2 = k > 1$ and  $H = 0$ the solutions under consideration
  obeying restrictions (\ref{3.3}) are absent. Indeed, using relations (\ref{3.4E}), 
  (\ref{3.4Q}) and  (\ref{3.4L}), we get 
  (see  (\ref{3.R1.0}), (\ref{3.R1.1}) and (\ref{3.R1.2})) 
   $\alpha > 0$,   
    \begin{equation}
     \alpha \Lambda = \frac{k}{8(k-1)}
      = \lambda_ {\infty }, 
    \label{3.R1.0a}
    \end{equation}
  and
  \begin{equation}
     h_1 = - h_2 = \pm \frac{1}{\sqrt{4 \alpha  (k - 1)}}.
     \label{3.R1.1a}
   \end{equation}
   We obtain $S_1 = k_1h_1 + k_2 h_2 = 0$, which is in contradiction 
   with our restriction $S_1 \neq 0$. Nevertheless, it may be verified that 
   the Hubble-like parameters 
   $H=0$ and $h_1$,  $h_2$ from (\ref{3.R1.1a})  obey the equations of motion 
   (\ref{2.3}), (\ref{2.4}) for $\alpha > 0$ and  $\Lambda$ from (\ref{3.R1.0a}). 
   This means that  we are led to a special solution, belonging
   to a subclass of solutions obeying $S_1 = 0$, which is out consideration in this paper.

 \section{The analysis of stability}

Here we study the stability of the solutions under consideration by using the
results of refs. \cite{ErIvKob-16,Ivas-16,ErIv-17-2}.

We put  the  restriction 
\begin{equation}
  \det (L_{ij}(v)) \neq 0
  \label{5.2}
\end{equation}
on the matrix 
\begin{equation}
L =(L_{ij}(v)) = (2 G_{ij} - 4 \alpha G_{ijks} v^k v^s).
   \label{5.1b}
 \end{equation}

We remind that for general cosmological setup with the metric 
\begin{equation}
 g= - dt \otimes dt + \sum_{i=1}^{n} e^{2\beta^i(t)}  dy^i \otimes dy^i,
 \label{5.3}
\end{equation}
we have  the  set of  equations \cite{ErIvKob-16} 
\begin{eqnarray}
     E = G_{ij} h^i h^j + 2 \Lambda  - \alpha G_{ijkl} h^i h^j h^k h^l = 0,
         \label{5.3.1} \\
         Y_i =  \frac{d L_i}{dt}  +  (\sum_{j=1}^n h^j) L_i -
                 \frac{2}{3} (G_{sj} h^s h^j -  4 \Lambda) = 0,
                     \label{5.3.2a}
          \end{eqnarray}
where $h^i = \dot{\beta}^i$,           
 \begin{equation}
  L_i = L_i(h) = 2  G_{ij} h^j
       - \frac{4}{3} \alpha  G_{ijkl}  h^j h^k h^l  
       \label{5.3.3},
 \end{equation}
 $i = 1,\ldots, n$.

Due to results of Ref. \cite{Ivas-16}  a fixed point solution
$(h^i(t)) = (v^i)$ ($i = 1, \dots, n$; $n >3$) to eqs. (\ref{5.3.1}), (\ref{5.3.2a})
obeying restrictions  (\ref{5.2}) is  stable under perturbations
\begin{equation}
 h^i(t) = v^i +  \delta h^i(t), 
\label{5.3h}
\end{equation}
 $i = 1,\ldots, n$,  as $t \to + \infty$, if (and only if)
 \begin{equation}
   S_1(v) = \sum_{i = 1}^{n} v^i >0
   \label{5.1s}
 \end{equation}
 and it is unstable if (and only if)
 \begin{equation}
    S_1(v) = \sum_{i = 1}^{n} v^i < 0.
    \label{5.1ns}
  \end{equation}
 
 In order to study the stability of solutions  we should verify the relation (\ref{5.2})
 for the solutions under consideration.  This verification was done (in fact)  in Ref.  \cite{ErIv-17-2}. 
 The proof of Ref.  \cite{ErIv-17-2}
 is based on  first three relations in (\ref{3.3}) and inequalities $k_1 > 1$, $k_2 > 1$ and $m >1$. We note the relation (\ref{3.5b}) was also used in this proof.
  
  Thus,  the any solution under consideration is stable 
  when relation (\ref{5.1s}) is obeyed while it is unstable when
  relation (\ref{5.1ns}) is satified. 
  
   Let us consider the case  $1 < m < k_1 < k_2$. For $H > 0$
   the relation (\ref{5.1s})   reads as
    \begin{equation}
       m  + k_1 x_1 + k_2 x_2 = 1  +  x_1 + x_2 >0
      \label{5.1xs}
    \end{equation} 
    or, equivalently,  
    \begin{equation}
            x_1 > X_4 = \frac{m - k_2}{k_2 - k_1}.
          \label{5.1xxs}
     \end{equation}
   Here the equation (\ref{3.15}) was used. For $H < 0$
   the stability condition  (\ref{5.1s}) reads 
   as  
   \begin{equation}
          m  + k_1 x_1 + k_2 x_2 = 1  +  x_1 + x_2 < 0,
         \label{5.1xsa}
       \end{equation} 
       or, equivalently, as 
      \begin{equation}
                 x_1 < X_4.
               \label{5.1xxsa}
          \end{equation}   

The non-stability condition (\ref{5.1ns}) reads as 
(\ref{5.1xxsa}) for $H > 0$ and as (\ref{5.1xxs})
for $H < 0$.

{\bf Proposition 3.} {\em  The solution to equations
  (\ref{2.3}), (\ref{2.4})  for  ansatz  (\ref{3.1}) with   $1 <  k_1 < k_2$,
  obeying the inequalities $H \neq 0$,  $H \neq h_1$, $H \neq h_2$, $h_1 \neq h_2$, 
  $S_1 = m H + k_1 h_1 + k_2 h_2 \neq 0$ is stable if and only if 
   $H (x_1 - X_4) > 0$ ($ X_4 = \frac{m - k_2}{k_2 - k_1}$) and it is unstable 
   if and only if $H (x_1 - X_4) < 0$. } 

Now we consider the case $H \neq 0$, $1 < m$, $1 < k_1 = k_2 = k$,  $m \neq k$. The exact 
solutions obtained in this section  obey first three relations in (\ref{3.3}) 
(since $x_1 \neq 1$, $x_2 \neq 1$ and 
$x_1 \neq x_2$) and hence  the key restriction (\ref{5.2}) is satisfied.

 The stability condition (\ref{5.1s}) in this case reads as, 
    \begin{equation}
        H ( m  + k_1 x_1 + k_2 x_2) = H (1  +  x_1 + x_2) = 
        H \left(1 - \frac{m - 1}{k - 1}\right) > 0,
        \label{5.2xs}
    \end{equation} 
see (\ref{4.25}), or, equivalently,  
 \begin{equation}
        H (k - m) > 0.
        \label{5.2xs}
    \end{equation}

 The non-stability condition (\ref{5.1ns}) reads as
  \begin{equation}
         H (k - m) < 0.
         \label{5.2nxs}
     \end{equation} 

Thus, we are led to the proposition.

{\bf Proposition 4.} {\em  The solution to equations
  (\ref{2.3}), (\ref{2.4})  for  ansatz  (\ref{3.1}) with   $1 < m$, $1 < k_1 = k_2 = k$,
  $m \neq k$, 
  obeying the inequalities $H \neq 0$,  $H \neq h_1$, $H \neq h_2$, $h_1 \neq h_2$, 
  $S_1 = m H + k h_1 + k h_2 \neq 0$ is stable if and only if 
   $ H (k - m) > 0$  and it is unstable 
      if and only if  $H (k - m) < 0$. } 

For $H > 0$ (or $\varepsilon_0 =  1$, see (\ref{4.5})) our special solutions  are 
stable for $k > m$ and they are unstable for $k < m$.
For $H < 0$ (or $\varepsilon_0 = - 1$) the solutions  are 
stable for $k < m$ and they are unstable for $k > m$.

{\bf The case $H = 0$.} Let us consider the solutions with $H=0$ and  $h_1$, $h_2$ 
from (\ref{3.R1.1}), (\ref{3.R1.2}), which are valid for $k_1 \neq k_2$, 
  $\alpha > 0$ and  $\Lambda$ from (\ref{3.R1.0}). Here $k_1 > 1$ and $k_2 > 1$. We obtain 
  \begin{equation}
    S_1 = k_1 h_1 + k_2 h_2 =  \pm (k_2 - k_1)
    \left(2 \alpha  (k_1 -1)(k_2 -1) (k_1 + k_2 - 2) \right)^{-1/2},
    \label{5.S1}  
 \end{equation}
 where $\pm$ is sign parameter in (\ref{3.R1.1}), (\ref{3.R1.2}).
 It follows from our analysis above that the solution with  $\pm (k_2 - k_1) > 0$ is stable. 
 This takes place when either  $k_2 > k_1$ and the sign $``+"$ is chosen in (\ref{3.R1.1}) and (\ref{3.R1.2}), 
 or if $k_2 < k_1$ and the sign $``-"$ is selected. 
  For  $\pm (k_2 - k_1) < 0$ the solution is unstable.  Here the restriction $m > 1$
  (which is used for the proof of  (\ref{5.2})) is also assumed.

\section{Solutions corresponding to zero variation of $G$}

Here we consider the special  solutions to equations 
(\ref{3.4E}), (\ref{3.4Q}), (\ref{3.4L})
with $H>0$, $3 < m < k_1 < k_2$ \cite{ErIv-17-2} (for $m =3$ see \cite{ErIv-19-2})
 \begin{equation}
   h_1 = \frac{m + 2 k_2 - 3}{k_2 - k_1} H, \qquad  h_2 = \frac{m + 2 k_1 - 3}{k_1 - k_2} H. 
    \label{6.1}
 \end{equation}
  
Here 
\begin{equation}
H   =   |k_1 - k_2| (- 2 \alpha P)^{-1/2},  
\label{6.2}
\end{equation}
$\alpha > 0$,
\begin{eqnarray}
P  =  P(m,k_1,k_2)  =- (m + k_1 + k_2 -3)(m (k_1 + k_2 -2) +
\nonumber \\
 k_1 ( 2 k_2 -5 ) + k_2 ( 2 k_1 -5 ) + 6) < 0, 
 \label{6.3} 
\end{eqnarray}
and 
\begin{equation}
\Lambda =  \Lambda(m,k_1,k_2),  
\label{6.2a}
\end{equation}
where
\begin{eqnarray}
    \Lambda(m,k_1,k_2) = \frac{1}{8 \alpha P^2} (m + k_1 + k_2 -3) [(k_1 + k_2)(k_1 + k_2 - 2)m^3 
   \qquad \qquad \qquad \qquad \qquad  
   \nonumber \\ 
   + (k_1^3 + k_2^3 + 11 (k_1^2 k_2 + k_1 k_2^2) - 19 (k_1^2 + k_2^2) - 22 k_1 k_2 + 18 (k_1 + k_2)) m^2 
   \qquad \qquad  \qquad \qquad \qquad
  \nonumber \\ -(8((k_1^3 + k_2^3)  - 63 (k_1 + k_2)^2 - 8 k_1^2 (k_1 - 11) k_2 
   \qquad \qquad \qquad \qquad  \qquad  
   \nonumber \\ 
   -  8 k_2^2 (k_2 - 11) k_1)  - 32 k_1^2 k_2^2 + 54 (k_1 + k_2)) m    \qquad \qquad \qquad \qquad \qquad
   \nonumber \\ 
    - ( 9 (k_1^3 + k_2^3) + 45 (k_1^2 + k_2^2) - 54 (k_1 + k_2) + 8 (k_1^2 + k_2^2) k_1 k_2 
    \qquad \qquad  \qquad \qquad \qquad \nonumber \\  \nonumber
    - 16 (k_1 + k_2  -10) k_1^2 k_2^2 - 9 (21 k_1 + 21 k_2  - 26) k_1 k_2 ) ]. \qquad 
       \qquad  \qquad    \qquad \qquad 
      % \label{6.4}
    \end{eqnarray}

These solutions describe accelerated exponential expansion of ``our'' $3d$ subspace  and
constant internal space volume factor, or zero variation of the effective gravitational 
constant (in Jordan frame) obeying the most stringent limitation
 on $G$-dot obtained by the set of ephemerides \cite{Pitjeva}, 
 when the following splitting of the Hubble-like parameters 
is keeping in mind:   
\begin{equation}
  \label{6.5}
   v =(\underbrace{H,H,H}_{``our'' \ space},\underbrace{\overbrace{H, \ldots, H}^{m-3}, 
   \overbrace{h_1, \ldots, h_1}^{k_1}, \overbrace{h_2, \ldots, h_2}^{k_2}}_{internal \ space}).
\end{equation}

It follows from Proposition 1 that  $\Lambda(m,k_1,k_2) > 0$.
Moreover, in this case we have 
 \begin{equation}
  x_1 =  \frac{m + 2 k_2 - 3}{k_2 - k_1} > 1. 
   \label{6.6}
 \end{equation}
Due to graphical analysis from  Sections 3 we get from  (\ref{6.6}) 
the following bounds  
 \begin{equation}
   0 < \lambda_{1}(m,k_1,k_2) < \Lambda(m,k_1,k_2) \alpha  <  \lambda_{\infty}(m,k_1,k_2)    
   \label{6.7}
 \end{equation}
for all $3 < m < k_1 < k_2$. 
 
 {\bf Remark.} {\em  It may be also shown that
  the effective gravitational  constant $G$ (in Jordan frame),
  calculated for our solutions, obeys  the limitation on $G$-dot from Ref. \cite{Pitjeva}, 
  when $\Lambda$ belongs to some vicinity of  $\Lambda(m,k_1,k_2)$, i.e. 
  $|\Lambda - \Lambda(m,k_1,k_2)| < \delta$ for some (small enough) $\delta > 0$.}

\section{Hubble-like parameters vs. constants of the model}

The initial contants of the model  are
 $\alpha_1 \neq 0$, $\alpha_2 \neq 0$ and  $\Lambda$.
 The solutions for Hubble-like parameters $H \neq 0$, $h_1$ and $h_2$, 
 which were analyzed above, depend upon  $\alpha = \alpha_2/\alpha_1 > 0$ and  $\lambda =  \Lambda  \alpha$.
 In this section we consider for simplicity the generic case $H \neq 0$.  
 The parameter $\alpha$ has the dimension of $L^2$ ($L$ is a length), while $\lambda$ is dimensionless one. 

Here we discuss the existence of certain combinations of Hubble-like parameters,
which either do not depend upon the parameters (or constants) of the model, i.e.
$\alpha$ and  $\lambda$, or depend only upon one of these constants.
Such combinations (or functions) of $H \neq 0$, $h_1$ and $h_2$ do exist.

Indeed, it follows from (\ref{3.4L}) that the Hubble-like parameters
for the solutions under consideration obey the following identity 
 \begin{equation} 
  \varphi_1 (H, h_1,h_2) \equiv 
  (m -1)  H + (k_1 - 1) h_1 +  (k_2 -1) h_2  = 0,  \label{7.1}
\end{equation}
$m > 1$, $k_1 > 1$ and $k_2 > 1$.
This is the first basic relation (of this section). 
By using  (\ref{3.9}) and (\ref{3.10b}) we get the second basic relation
 \begin{equation}
 \varphi_2 (H, h_1,h_2) \equiv  
 (m -1)  H^2 + (k_1 - 1) h_1^2 +  (k_2 -1) h_2^2  =  \frac{1}{ 2 \alpha}.       \label{7.2}
  \end{equation} 
The third basic relation is just  (\ref{3.13a}) which we rewrite here as
 \begin{equation}
   \varphi_3 (H, h_1,h_2) \equiv  f(h_1/H) = \lambda,   \label{7.3}
 \end{equation} 
where $f(x_1)$ is the rational function defined in (\ref{3.13a}).

In the $3d$ space of  Hubble-like parameters $H, h_1, h_2$, relation 
 (\ref{7.1}) describes a plane while  (\ref{7.2}) 
corresponds to an ellipsoid. The intersection of this plane  and ellipsoid 
gives us an ellipse ${\cal E}$. For   $m=3$,  $k_1=4$,  $k_2 = 5$ and $\alpha =1$  this 
intersection is depicted at Figure 6. For $H \neq 0$ and $m < k_1 < k_2$ 
 the solutions  for  $(H, h_1, h_2)$ are described
 by  $1$-dimensional manifold   ${\cal E}_{sol} = {\cal E}
  \setminus \{ N, S, Y_1, Y_2, Y_3, Y_4, - Y_1, - Y_2,- Y_3, -Y_4 \} $,
 where points $N, S$ correspond to $H = 0$, points  $Y_1, Y_2, Y_3, Y_4$ 
 correspond to $H >0$ and relations  $h_1/H = X_1$, $h_2/H = X_2$, 
  $h_3/H_3 = X_3$, $h_4/H_4 = X_4$, respectively 
 (see (\ref{3.x1}), (\ref{3.x2}), (\ref{3.x3}), (\ref{3.x4})).
  Thus, the manifold ${\cal E}_{sol}$ is an  $1$-dimensional manifold, which
  is obtained from the ellipse ${\cal E}$ by deleting $10$ points. 
  It is a disjoint union of ten arcs. Any of these arcs
  is parametrized by the pair $(\lambda, s)$, where 
  $s$ is the number of the arc and 
  $\lambda$ is local coordinate given by  (\ref{7.3}). 
  %In this case
  %relation (\ref{7.3}) defines an atlas of  $10$ charts on ${\cal E}_{sol}$ .
  Analogous consideration may be done for the case $m \neq k_1 = k_2$:
  in this case one should delete $8$ points from ${\cal E}$  to obtain ${\cal E}_{sol}$.

 \begin{figure}
	\begin{center}
	\includegraphics[width=0.75\linewidth]{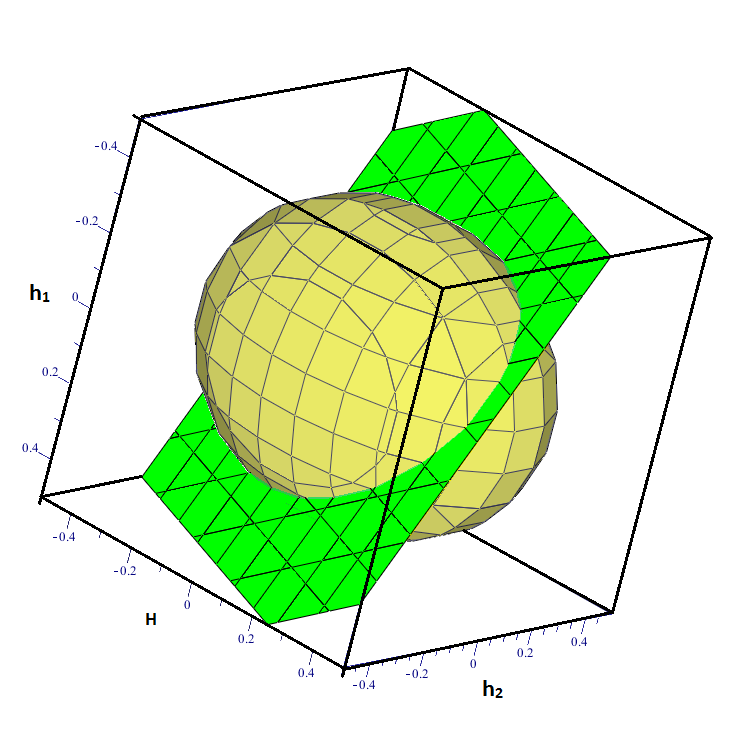}
		\caption{The graphical representation (in Hubble-like variables $H, h_1, h_2$) of intersection 
		of plane (see (\ref{7.1})) and ellipsoid (see (\ref{7.2})) for  $m=3$,  $k_1=4$,  $k_2 = 5$ and 
		$\alpha = 1$.}
	\label{rfig:6}
	\end{center}
 \end{figure}

 It should be noted that (\ref{7.1}) implies the following identity for scale factors
 $a_i(t) = \exp(h_i t + \beta_i)$, $i = 0,1,2$, ($h_0 = H$)
 \begin{equation} 
    (a_0(t))^{m-1} (a_1(t))^{k_1 -1} (a_2(t))^{k_2 - 1} = 
    {\rm  const },                                               \label{7.4}
 \end{equation}
 or 
\begin{equation} 
    v(t) = (a_0(t))^{m} (a_1(t))^{k_1} (a_2(t))^{k_2} = 
    {\rm const } \times  a_0(t) a_1(t) a_2(t).                 \label{7.5}
 \end{equation}
Here $v(t) = \exp(\sum_{i = 0}^{2} (h_i t + \beta_i))$ is volume scale factor which is
(exponentiallly) increasing in time for stable solutions (with $H + h_1 + h_2 > 0$) and 
decreasing in time for unstable ones (with $H + h_1 + h_2 < 0$).

\section{Conclusions}

We have considered the  $D$-dimensional  Einstein-Gauss-Bonnet (EGB) model with a $\Lambda$-term 
(or EGB$\Lambda$ model) and two (non-zero)
constants   $\alpha_1$ and $\alpha_2$.  The metric was chosen to be diagonal ``cosmological'' one. 
Here we were dealing (mainly)
 with a class of solutions with  exponential time dependence of three scale factors, governed by three non-coinciding
 Hubble-like parameters $H \neq 0$, $h_1$ and $h_2$, corresponding to factor spaces of dimensions $m > 1$,  $k_1 > 1$
  and $k_2 > 1$, respectively, with the restriction imposed:  $S_1 = m H + k_1 h_1 + k_2 h_2 \neq 0$,  and 
  $D = 1 + m + k_1 + k_2$.  

We have studied the solutions in two cases: i) $ m < k_1 < k_2$ and ii)  $1< k_1 = k_2 = k \neq m$. (The solutions under consideration with $k_1 = k_2 = m$ are absent.) We have shown that in both cases the solutions  exist only if:  $\alpha = \alpha_2 / \alpha_1 > 0$,  $\lambda = \alpha \Lambda > 0$ and  the dimensionless parameter of  the model $\lambda$ obeys certain restrictions, e.g. upper and lower bounds depending upon  $m$, $k_1$ and $k_2$ 
(see Proposition 1).   In the case ii) we have found  explicit  exact solutions (see Proposition 2). 

Our consideration  used the  so-called  Chirkov-Pavluchenko-Toporensky splitting trick from 
Ref. \cite{ChPavTop1} (see also \cite{ErIv-17-2}) which allowed us to  reduce  the problem under consideration to  master equation    $ \lambda = f(x_1)$ (\ref{3.13}), where  $x_1  = h_1/H$. This master equation  is equivalent to polynomial equation   (\ref{3.17}) for $x_1$ which is of fourth order (in generic case) or less depending upon  $\lambda$. Thus, the master equation may be solved in radicals for all  $m > 1$, $k_1 > 1$ and $k_2 > 1$.  Our restrictions on $\lambda$    were  obtained by analysing the equation    $ \lambda = f(x_1)$ with the use of the formulas  for the derivative $df/dx_1$, i.e. (\ref{3.f}) and    (\ref{3.ff})  in  cases  i) and ii), respectively. %Here we did not  use the algebraic approach to polynomial equations   of  fourth  or third order. 
%(It looks that the analytical way is more  economic and effective for our task of cosmological origin.)
 In the case i)  $m < k_1 < k_2$ the  extremum  points of the function $ f(x_1)$ are just  four non-coinciding points: $X_1, X_2, X_3, X_4$ (see (\ref{3.x1}),  (\ref{3.x2}),  (\ref{3.x3}), (\ref{3.x4})) which   are  exactly four values of $x_1$  forbidden by restrictions   $ H \neq h_1$, $ H \neq h_2$,  $h_1 \neq h_2$, $S_1 = m H + k_1 h_1 + k_2 h_2 \neq 0$, respectively.
In the case ii) $1 < k_1 = k_2 \neq m $ we have three forbidden  points: $X_1, X_2, X_3$.

The stability of the solutions (as  $t \to + \infty$) in a class of cosmological solutions with diagonal  metrics was analyzed for both cases ((i) and (ii)) and subclasses of stable and non-stable  solutions were singled out. We have proved that  in the case i) the solutions with $H > 0$ are stable for   
$x_1 = h_1/H  > X_4 = \frac{m - k_2}{k_2 - k_1}$ and unstable for $x_1 < X_4$ (see Proposition 3). It was proved  that in the case ii) the solutions with $H > 0$  are stable for $k > m$ and unstable for $k < m$ (see Proposition 4).   The stability conditions for $H < 0$ are equivalent to instability conditions for $H >0$ and vice versa.  
The solutions of first class  i) contains a  subclass of stable solutions  describing an exponential expansion of  $3$-dimensional subspace with Hubble-like parameter $H > 0$ and zero variation of the effective gravitational constant $G$ (in the Jordan frame) \cite{ErIv-17-2} (see Section 6). 

Some of the results obtained in this paper may be considered as non-trivial and  unexpected ones. 
Indeed,  let us compare   the solutions  governed by three different Hubble-like parameters $H > 0$, $h_1$,  $h_2$  with  the  solutions from Ref. \cite{IvKob-18-2} obtained for two non-coinciding Hubble-like parameters $H > 0$ and $h$ corresponding to factor spaces of dimensions $m > 2$ and $l > 2$ with $mH +  l h \neq 0$.   
Here we have found  that our solutions take place only for $\alpha  > 0$ and $\Lambda >0$, while in the case of  Ref. \cite{IvKob-18-2} we have two branches with (a) $\alpha > 0$, $ - \infty <  \Lambda \alpha < \lambda_{+}(m,l)$ and  (b) $\alpha < 0$, $  \Lambda |\alpha| > \lambda_{-}(m,l)$, where    $ \lambda_{\pm}(m,l) > 0$. The solutions from    Ref. \cite{IvKob-18-2} with $\alpha > 0$  exist for any $\Lambda  \in ( - \infty, 0 ] $, while in our case  such solutions are absent. We note that the absence of solutions for $\Lambda = 0$ may be considered as a special non-trivial result. For two different Hubble parameters such solutions (with $\Lambda = 0$ and $\alpha > 0$) were described in Ref. \cite{IvKob-19-EPJC}.   As it is proved here,  in the  case  of three Hubble-like parameters 
(with the restrictions imposed above) the allowed gap for $\Lambda$ is bounded (at the top and the bottom).   

 Here we have also considered (for a completeness) the case $H =0$ and have found that  the solutions exist only for $k_1 \neq k_2$, $\alpha > 0$ and fixed value of $\Lambda > 0$  from (\ref{3.R1.0}).  In this case we have two   opposite in sign  solutions
 for $(h_1, h_2)$ with  one solution being stable and the second one - unstable. 

For possible physical (e.g. cosmological) applications one may keep in mind  a dimensional reduction  of the model under consideration to $d=4$ which lead us to $4d$ Horndeski type model with a set of scalar fields. In this case one will obtain   $(1+3)$-dimensional inflationary (cosmological) solution with Hubble parameter $H >0$ and several  scalar fields (coming from scale factors) with linear dependence upon the time variable (governed by $h_1$ and $h_2$). The effective cosmological term $ \Lambda_0 = 3H^2$  will have a nontrivial dependence  upon the  ``bare'' multidimensional cosmological constant $ \Lambda$, the dimensions of  factor spaces $m$, $k_1$ and $k_2$ and the parameter $\alpha$ (for any root of  polynomial equation for $x_1$).

\renewcommand{\theequation}{\Alph{section}.\arabic{equation}}
\renewcommand{\thesection}{\Alph{section}}
\setcounter{section}{0}

\section{Appendix}

Here we prove several technical lemmas.

{\bf Lemma 1}.
 {\em Let 
 \begin{equation}
 v(m,l,k) = (k+ l)m^2 + (m + l )k^2 +(m + k)l^2 - 6mlk,    \label{A.1}        
 \end{equation}
 where   $m,l,k$  are natural  numbers. Then   $v(m,l,k) = 0$ only if 
 $m = l = k$; in other cases  $v(m,l,k) > 0$}.

{\bf Proof}. Since the $v(m,l,k)$ is symmetric in variables we put 
without loss of generality $m \geq l \geq k$. We have   
    $m = k+p+q$,  $l = k+p$, where  $p \geq 0$ and $q \geq 0$.
We get 
 \begin{equation}
 v = v(m,l,k) = (2p^2+2qp+2q^2)k+2p^3+3qp^2+ p q^2.  \label{A.2}
 \end{equation}
For   $p = q =  0$  ($m=k=l$) we have $v = 0$.
For $p > 0$, $q > 0$ we have $v > 0$. If $p =  0$ ($k = l$) 
and $q > 0$ ($m > l$) we get $v = 2q^2 k > 0$ for $k >1$. 
 For $q =  0$ ($m = l$) and $p > 0$ ($l >k$) we find $v =  2p^2 k  +2 p^3 > 0$.
  The lemma is proved. 

{\bf Lemma 2.}
 {\em Let 
\begin{eqnarray}
w(m,l,k) = (k+ l - 2)m^2 + (m + l - 2)k^2 +(m + k - 2)l^2
\nonumber \\
   + 2km + 2km + 2lk   -  6mlk, \label{A.3}
\end{eqnarray}              
where   $m,l,k$ are natural  numbers non equal to $1$. Then   $w(m,l,k) = 0$ only if 
$m = l = k$. In other cases  $w(m,l,k) > 0$.}

{\bf Proof.} Since the $w(m,l,k)$ is symmetric in variables we put 
without loss of generality $m \geq l \geq k$. We have   
  $m = k+p+q$,  $l = k+p$, where  $p \geq 0$ and $q \geq 0$.
We get 
  \begin{equation}
 w = w(m,l,k) = (2p^2 + 2qp + 2q^2)(k -1) + 2p^3+ 3qp^2 + q^2p.  \label{A.4}
  \end{equation}
For   $p = q =  0$  ($m=k=l$) we have $w = 0$.   
For $p > 0$, $q > 0$ we have $w > 0$ (for all $k$). If $p =  0$ ($k = l$) and $q > 0$ 
($m >  l$) we get $w = 2q^2(k -1) > 0$ for $k >1$. 
 For $q =  0$ ($m=l$) and $p > 0$ ($l > k$) we find $w =  2p^2 (k -1)  +2 p^3 > 0$. 
 The lemma is proved. 

{\bf Lemma 3.}
 {\em For all $1 < m < k_1 < k_2$
\begin{equation}
z_1 = (2k_1-k_2 - 1)m - 2k_1k_2 -  4k_1  + 5k_2 -  k_2^2  + 2k_1^2 < 0. 
    \label{A.5}
\end{equation} 
}
{\bf Proof.} Let us denote 
\begin{equation}
k_1 = m + 1 + y_1, \qquad k_2 = k_1 + 1 + y_2.
 \label{A.6}
\end{equation}
Due to $m < k_1 < k_2$ we get $y_1 \geq 0$ and $y_2 \geq 0$.      
The substitution of  (\ref{A.5}) into $z_1$ gives us 
\begin{equation}
z_1 = -y_2^2 + (-4y_1-5m-1)y_2- y_1^2+(-m-5)y_1 - 6m \leq -6m < 0. 
\label{A.7}
\end{equation}
The lemma is proved.

{\bf Lemma 4.}
 {\em For all $1 < m < k_1 < k_2$
\begin{eqnarray}
 z_3 = - (k_1 - k_2)^2 m^2 - ((k_1^2 - 6k_1 +6)k_1 + (k_2^2 - 6k_2 +6)k_2 \nonumber \\
  + (k_1 + k_2)k_1k_2 - 4)m -  2(k_1 + k_2) \nonumber \\
  +  (12 - 6k_1 - 6k_2  + (k_1 + k_2)^2 )k_1k_2 > 0.
  \label{A.8}
  \end{eqnarray} 
}
{\bf Proof.}   
Substituting  (\ref{A.6}) into $z_3$ we obtain 
\begin{eqnarray}
z_3 = (y_1+1)y_2^3+(5y_1^2+(6m+7)y_1+6m+2)y_2^2+
\nonumber \\
(8y_1^3+(18 m +16)y_1^2 + (12m^2+24m+11)y_1+2 m^3+6 m^2+12m+1)y_2
 \nonumber \\
+ 4 y_1^4+(12 m+12) y_1^3+ (12 m^2+30m+11)y_1^2+
 \nonumber \\
 (4m^3 + 24m^2 + 18m + 5)y_1
 + 6 m^3 + 6 m^2 + 6m \geq 
 \nonumber \\ 
  6 m^3 + 6 m^2 + 6m > 0, \qquad
 \label{A.9}
\end{eqnarray}
since $y_1 \geq 0$ and $y_2 \geq 0$.
The lemma is proved.

%\begin{center}
 {\bf Acknowledgments}
%\end{center}

The publication has been prepared with the support of the ``RUDN University Program 5-100'' (recipient V.D.I., mathematical model development). The reported study was funded by RFBR, project number 19-02-00346 
(recipient K.K.E., simulation model development).
%\newpage


\small

\begin{thebibliography}{99}


\bibitem{Zwiebach}
 B. Zwiebach, Curvature squared terms and string theories,
  Phys. Lett.  B {\bf 156}, 315 (1985).

 \bibitem{FrTs1}
 E.S. Fradkin and A.A. Tseytlin, Effective field theory from quantized strings,
  Phys. Lett. B {\bf 158}, 316-322 (1985).

 \bibitem{FrTs2}
 E.S. Fradkin and A.A. Tseytlin, Effective action approach to superstring
 theory,  Phys. Lett. B {\bf   160}, 69-76 (1985).

 \bibitem{GW}
 D. Gross and E. Witten,
 Superstrings modifications of  Einstein's equations,
 Nucl. Phys. B {\bf 277},  1 (1986).

  
\bibitem{Ishihara}
 H. Ishihara, Cosmological solutions of the extended Einstein
  gravity with the Gauss-Bonnet term,
  Phys. Lett. B {\bf  179}, 217 (1986).

 \bibitem{Deruelle}
 N. Deruelle, On the approach to the cosmological
 singularity in quadratic theories of gravity: the Kasner
 regimes,   Nucl. Phys. B  {\bf  327},  253-266 (1989).

\bibitem{NojOd0}
S. Nojiri and S.D. Odintsov, Introduction to modified gravity and
gravitational alternative for Dark Energy,
 Int. J. Geom. Meth. Mod. Phys. {\bf  4}, 115-146 (2007);
   hep-th/0601213.

\bibitem{CElOdZ}
G. Cognola, E. Elizalde, S. Nojiri, S.D. Odintsov and S.
Zerbini, One-loop effective action for non-local modified
Gauss-Bonnet gravity in de Sitter space,
 Eur. Phys. J. C  {\bf   64}(3), 483-494 (2009);
 arXiv: 0905.0543.

\bibitem{ElMakObOsFil}
 E. Elizalde, A.N. Makarenko, V.V. Obukhov, K.E. Osetrin  and
 A.E. Filippov,  Stationary vs. singular points in an accelerating
 FRW cosmology derived from six-dimensional Einstein-Gauss-Bonnet
 gravity,  Phys. Lett. B {\bf  644}, 1-6 (2007);  hep-th/0611213.

  \bibitem{BambaGuoOhta}
 K. Bamba, Z.-K. Guo and N. Ohta, Accelerating Cosmologies in the
 Einstein-Gauss-Bonnet theory with dilaton, Prog. Theor. Phys.
 {\bf 118},  879-892 (2007); arXiv: 0707.4334.

 \bibitem{TT}
 A. Toporensky and P. Tretyakov,
 Power-law anisotropic cosmological solution in 5+1 dimensional
 Gauss-Bonnet gravity,  Grav. Cosmol. {\bf 13}, 207-210 (2007);
  arXiv: 0705.1346.

 \bibitem{PTop}
 S.A. Pavluchenko and  A.V. Toporensky, A note on differences
 between $(4+1)$- and $(5+1)$-dimensional anisotropic cosmology in the
 presence of the Gauss-Bonnet term, Mod. Phys. Lett.  A {\bf 24},
  513-521 (2009).

\bibitem{KirMak}
I.V. Kirnos and A.N. Makarenko,
Accelerating cosmologies in Lovelock gravity with dilaton,
 Open Astron. J. {\bf 3}, 37-48 (2010);  arXiv: 0903.0083.

\bibitem{Pavl}
 S.A. Pavluchenko,
 On the general features of Bianchi-I cosmological models in
 Lovelock gravity,  Phys. Rev. D {\bf  80}, 107501 (2009);
 arXiv: 0906.0141.

 \bibitem{KirMPTop}
 I.V. Kirnos, A.N. Makarenko, S.A. Pavluchenko and A.V. Toporensky,
 The nature of singularity in multidimensional anisotropic
 Gauss-Bonnet cosmology with a perfect fluid,
  Gen. Rel. Grav. {\bf 42}, 2633-2641 (2010);
  arXiv: 0906.0140.

\bibitem{Iv-09}
  V.D. Ivashchuk,
  On anisotropic Gauss-Bonnet cosmologies in (n + 1) dimensions,
  governed by an n-dimensional Finslerian 4-metric,  Grav. Cosmol.
  {\bf 16}(2), 118-125 (2010); arXiv: 0909.5462.

 \bibitem{Iv-10}
 V.D. Ivashchuk,  On cosmological-type solutions in
 multidimensional model with  Gauss-Bonnet term,
 Int. J. Geom. Meth. Mod. Phys.
 {\bf 7}(5), 797-819 (2010);  arXiv: 0910.3426.

\bibitem{MaOh}
 K.-i. Maeda and N. Ohta,
 Cosmic acceleration with a negative cosmological constant in
 higher dimensions,  JHEP {\bf 1406}: 095 (2014); arXiv:1404.0561.

\bibitem{ChPavTop}
D. Chirkov, S. Pavluchenko and A. Toporensky, Exact exponential solutions
in Einstein-Gauss-Bonnet flat anisotropic cosmology,
 Mod. Phys. Lett. A {\bf  29},  1450093 (11 pages) (2014);  arXiv:1401.2962.

\bibitem{ChPavTop1}
D. Chirkov, S.A. Pavluchenko and A. Toporensky,
Non-constant volume exponential solutions in higher-dimensional
Lovelock cosmologies,  Gen. Rel. Grav.  {\bf 47}: 137 (33 pages) (2015); arXiv: 1501.04360.


 \bibitem{Pavl-15}
 S.A. Pavluchenko, Stability analysis of exponential solutions in Lovelock cosmologies,
 Phys. Rev. D {\bf 92}, 104017 (2015); arXiv: 1507.01871.

\bibitem{Pavl-16}
 S.A. Pavluchenko, Cosmological dynamics of spatially flat Einstein-Gauss-Bonnet models in various dimensions: Low-dimensional  $\Lambda$-term case, Phys. Rev. D {\bf 94}, 084019 (2016); arXiv: 1607.07347.  
    
\bibitem{ErIvKob-16}
 K.K. Ernazarov, V.D. Ivashchuk and A.A. Kobtsev,
 On exponential solutions in the Einstein-Gauss-Bonnet cosmology,
 stability and variation of G,  
   Grav.  Cosmol. {\bf 22} (3), 245-250 (2016).

\bibitem{CGPT}
F. Canfora, A. Giacomini, S.A. Pavluchenko and A. Toporensky,
Friedmann dynamics recovered from compactified Einstein-Gauss-Bonnet cosmology,
arXiv:1605.00041.

\bibitem{Ivas-16}
V.D. Ivashchuk, On stability of exponential cosmological solutions
with non-static volume factor in the Einstein-Gauss-Bonnet model,
 Eur. Phys. J.  C {\bf 76}, 431 (2016); arXiv: 1607.01244v2.

\bibitem{ErIv-17-2}
K.K. Ernazarov and V.D. Ivashchuk, Stable exponential cosmological solutions 
with zero variation of G and three different Hubble-like parameters in the 
Einstein-Gauss-Bonnet model with a $\Lambda$-term, 
Eur. Phys. J. C 77: 402  (2017) (7 pages);  arXiv:1705.05456.

\bibitem{IvKob-18-2}
V.D. Ivashchuk and A.A. Kobtsev, Stable exponential cosmological solutions
 with two factor spaces in the Einstein-Gauss-Bonnet model with a $\Lambda$-term, 
 Gen. Rel. Grav., {\bf 50}, 119 (2018); arXiv:1712.09703v4. 

  
% % % % % % % % % % % % % % % % % % % % % % % % % %

\bibitem{NOO-17}
 S. Nojiri,  S.D. Odintsov and V.K. Oikonomou, 
Modified Gravity Theories on a Nutshell: Inflation, Bounce and Late-time Evolution,
  Phys. Rept., {\bf 692},  1-104 (2017).

\bibitem{BSCapAL-18}
 M. Benetti, S. Santos da Costa and S. Capozziello, J.S. Alcaniz and M. De Laurentis, 
Observational constraints on Gauss-Bonnet cosmology,
 Int. J.  Mod. Phys., {\bf 27}, 1850084 (2018).

\bibitem{NOO-19}
S. Nojiri, S.D. Odintsov and V.K. Oikonomou, 
Unifying Inflation with Early and Late-time Dark Energy in $F(R)$ Gravity;
 arXiv: 1912.13128. 

% % % % % % % % % % % % % % % % % % % % % % % % % % % % % % % % %

\bibitem{Riess}
 A.G. Riess  et al. Observational evidence from supernovae
 for an accelerating  universe and a cosmological constant,
  Astron. J. {\bf 116}, 1009-1038 (1998).


\bibitem{Perl}
 S. Perlmutter  et al., Measurements of Omega and Lambda from 42 High-Redshift  Supernovae.
  Astrophys.  J. {\bf 517},  565-586 (1999).

 \bibitem{Kowalski}
 M. Kowalski, D. Rubin et al., Improved cosmological constraints from new,
 old and combined supernova datasets,
  Astrophys.  J. {\bf 686} (2), 749-778 (2008); arXiv: 0804.4142.

%\bibitem{Ade}
%P.A.R. Ade  et al. [Planck Collaboration], Planck 2013 results.
%I. Overview of products and scientific results,
% Astron. Astrophys. {\bf 571}, A1 (2014); arXiv: 1303.5076.

% % % % % % % % % % % % % % new 2 % % % % % % % % % % %

\bibitem{Lovelock}
 D. Lovelock, The Einstein tensor and its generalizations, J. Math Phys. {\bf 1971},  12, 498.

%\bibitem{Chern}
%Chern S.-S. On the Curvatura Integra in a Riemannian Manifold.
%{\it Ann. Math.}, {\bf 1945}, {\it 46}, 674-684.

% % % % % % % % % % % % % % %

\bibitem{ErIv-19-GC}
K.K. Ernazarov and V.D. Ivashchuk, 
Examples of Stable Exponential Cosmological Solutions
with Three Factor Spaces in EGB Model with a $\Lambda$-Term,
 Grav.  Cosmol.  {\bf 25}, 164-168  (2019). 

\bibitem{ErIv-19-2}
K.K. Ernazarov and V.D. Ivashchuk,
Stable  exponential cosmological  solutions with three different Hubble-like parameters in 
 $(1 + 3 + k_1 + k_2)$-dimensional  EGB  model  with a $\Lambda$-term, Symmetry 
 {\bf  12} (2), 250 (24 pages) (2020). 

\bibitem{Pitjeva}
E.V. Pitjeva,
Updated IAA RAS Planetary Ephemerides-EPM2011
and Their Use in Scientific Research, Astron. Vestnik
{\bf 47} (5),  419-435 (2013), arXiv: 1308.6416.

\bibitem{IvKob-19-EPJC}
V.D. Ivashchuk and A.A.  Kobtsev, 
Exponential cosmological solutions with two factor spaces in EGB model 
with $\Lambda = 0$ revisited,   Eur. Phys. J. C {\bf 79}, 824 (2019).

\end{thebibliography}
\end{document}